\documentclass[aps,
twocolumn,
showpacs,preprintnumbers,amsmath,amssymb,superscriptaddress,prx,floatfix,10pt]{revtex4-2}
\usepackage{graphicx}
\usepackage{mathtools}
\usepackage{bbm}
\usepackage[normalem]{ulem}
\usepackage{xcolor}
\usepackage{comment}
\usepackage{amsthm} 
\usepackage{amssymb}   
\usepackage{cancel}
\usepackage[colorlinks, linkcolor=blue, citecolor=blue]{hyperref}

\renewcommand{\bf}[1]{\mathbf{#1}}

\begin{document}
\title{Odd relaxation in three-dimensional Fermi liquids}

\author{Seth Musser}
\email[]{swmusser@gmail.com}
\affiliation{Condensed Matter Theory Center and Joint Quantum Institute, Department of Physics, University of Maryland, College Park, Maryland 20742, USA}
\author{Sankar Das Sarma}
\affiliation{Condensed Matter Theory Center and Joint Quantum Institute, Department of Physics, University of Maryland, College Park, Maryland 20742, USA}
\author{Johannes Hofmann}
\affiliation{Department of Physics, Gothenburg University, 41296 Gothenburg, Sweden}
\affiliation{Nordita, Stockholm University and KTH Royal Institute of Technology, 10691 Stockholm, Sweden}

\begin{abstract}
Recent theoretical works predict a hierarchy of long-lived, non-hydrodynamic modes in two-dimensional Fermi liquids arising from the feature---supposedly unique to two dimensions---that relaxation by head-on scattering is not efficient in the presence of Pauli blocking. This leads to a parity-based separation of scattering rates, with odd-parity modes relaxing much more slowly than even-parity ones. In this work, we establish that a similar effect exists in isotropic three-dimensional (3D) Fermi liquids, even though relaxation does not proceed solely by head-on scattering. We show that while the relaxation rates of even and odd modes in 3D share the same leading-order $\sim T^2$ low-temperature scaling typical of Fermi liquids, their magnitudes differ, with odd-parity modes relaxing more slowly than even ones for a broad class of interactions. We find a relative difference between odd-parity and even-parity relaxation rates as large as~$40\%$ just by Pauli blocking alone, with a strong additional dependence on the scattering potential, such that the odd-even staggering is further enhanced by interactions that favor large-angle scattering. We identify signatures of these odd-parity relaxation rates in the static transverse conductivity as well as the transverse collective mode structure. Our results establish the unexpected existence of a tomographic like regime in higher-dimensional Fermi liquids and suggest experimental probes via transport measurements.
\end{abstract}

\maketitle

\section{Introduction}

Sixty years ago, Gurzhi~\textcite{gurzhi62,gurzhi68} identified the possibility of hydrodynamic behavior in a system of electrons. The necessary condition for hydrodynamic rather than the better-known diffusive transport is that the momentum-conserving inelastic electron-electron scattering is stronger than the momentum-conservation-breaking impurity (or phonon) scattering in the system, which we assume to be valid in our theory~\cite{ahn_hydrodynamics_2022}. In such hydrodynamic systems, including hydrodynamic Fermi liquids, microscopic collisions are assumed to quickly relax any inhomogeneity that is not linked to a conserved quantity, which is referred to as reaching ``local equilibrium." The remaining dynamics is that of conserved densities---particle number, momentum, and energy---which obey conservation equations and relax via spatial transport, described by hydrodynamic equations, to global equilibrium \cite{Tong_2025, lucas_hydrodynamics_2018, musser_observable_2024, fritz_hydrodynamic_2024, khveshchenko_strange_2024, khveshchenko_collective_2025}. The universal nature of these hydrodynamic equations has allowed for the observation of water like behavior in electron systems; bathtub like vortices are just one recent example~\cite{aharonsteinberg22, krebs_imaging_2023}.

\begin{figure}
        \includegraphics[width=0.45\columnwidth]{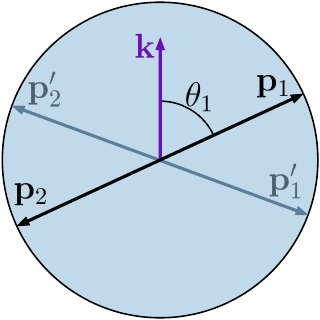}
        \includegraphics[width=0.45\columnwidth]{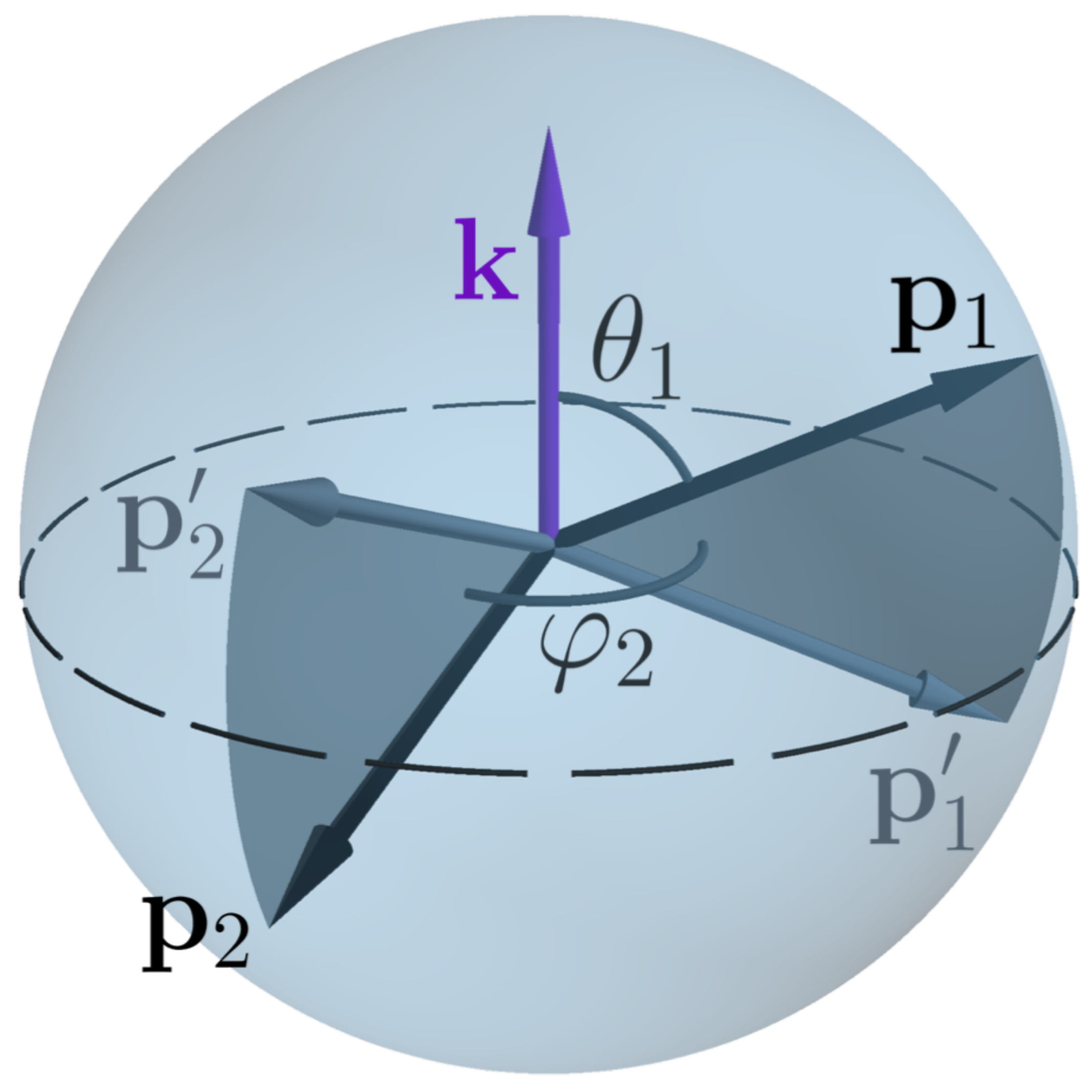}
    \caption{Momentum space picture of the dominant two-particle collision processes for an isotropic Fermi liquid at low temperatures in (a)~2D and (b)~3D. Shaded regions indicate the occupied states in equilibrium, and ${\bf p}_1$ and ${\bf p}_2$ indicate initial and ${\bf p}_1'$ and ${\bf p}_2'$ final momenta, where \mbox{${\bf k} = {\bf p}_1 - {\bf p}_1'$} is the momentum transfer in the collision. In both cases, Eq.~\eqref{eqn:angle_fixed} constrains the scattering momenta to form a rectangle. In 2D, this enforces head-on scattering with only a single free parameter $\theta_1$, the angle between $\bf{k}$ and $\bf{p}_1$, which leads to a parity-based separation of relaxation rates. In 3D, in contrast, an additional parameter  $\varphi_2$ enters that describes the angle between \mbox{$\bf{p}_1+\bf{p}_1'$} and \mbox{$\bf{p}_2+\bf{p}_2'$}. Unless \mbox{$\varphi_2 = \pi$}, head-on scattering will not occur. Interactions that enhance large-angle scattering will prefer \mbox{$\varphi_2\sim \pi$} and thus enhance the even-odd effect, while those suppressing it will reduce the even-odd effect.
    }
	\label{fig:2d_vs_3d}
\end{figure}

Over the last few years, however, a number of theoretical works have revealed a non hydrodynamic ``tomographic'' transport regime in two-dimensional (2D) Fermi liquids~\cite{ledwith_angular_2019,ledwith19b,hofmann_collective_2022,hofmann_anomalously_2023}. Such tomographic transport is only observable in Fermi liquids and is marked by the appearance of many long-lived modes that are not related to conserved quantities. The origin of this effect is linked to  microscopic constraints on the collisions that relax deviations from local equilibrium. While in any system, from water to Fermi liquids, the dominant collisions will be two-particle collisions, Fermi liquids are special because they obey Pauli exclusion. At low temperatures, this implies that dominant two-particle collisions in an isotropic 2D Fermi liquid are head-on scattering processes, i.e., two particles with Fermi momentum $\bf{p}_F$ and $-\bf{p}_F$ scattering to momenta $\bf{p}'_F$ and $-\bf{p}'_F$, where $\bf{p}_F$ and $\bf{p}_F'$ have arbitrary orientation on the Fermi surface~\cite{laikhtman_electron-electron_1992}. Such a process is displayed in Fig.~\ref{fig:2d_vs_3d}(a), where ${\bf p}_1$ and ${\bf p}_2$ are the initial and ${\bf p}_1'$ and ${\bf p}_2'$ are the final momenta, and the blue shaded area marks the occupied states in equilibrium. These head-on collision processes are able to relax deviations in the quasiparticle distribution from equilibrium that are symmetric under parity (i.e., with an excess or a depletion of quasiparticles at opposing points on the Fermi surface). The relaxation rates of these even parity modes scale at low temperatures as \mbox{$\sim T^2/\hbar T_F$}, where~$T_F$ is the Fermi temperature, which is the typical Fermi liquid scattering rate (note that the collective even- or odd-parity deformations are distinct from single-quasiparticle decay, which show a logarithmic-in-temperature correction---i.e., a scaling of $T^2\log T$ with $T^2$ being dominant---to the quasiparticle scattering rate in 2D~\cite{giuliani82,zheng96, liao_two-dimensional_2020}). Crucially, however, an inhomogeneity that is not symmetric under parity \textit{cannot} relax at this order, instead relaxing by subleading processes at a much smaller rate. This even-odd tomographic distinction is assumed to be specific to two-dimensional Fermi liquids~\cite{gurzhi_electron-electron_1995,ledwith_angular_2019}.

This picture has been made more precise in a full exact diagonalization of the electron collision integral~\cite{hofmann_anomalously_2023, nilsson_nonequilibrium_2024} to determine the precise properties of the odd-parity modes. Such analysis revealed a finite number of anomalously long-lived modes below \mbox{$T\lesssim0.15T_F$}, the number of which scales roughly as~${\it O}(\sqrt{T/T_F})$, with an asymptotic~\mbox{$\sim T^4/\hbar T_F^3$} low-temperature scaling of the relaxation rate. Each of the long-lived odd-parity modes is described by an odd angular harmonic index~$m$ that describes the angular deformation of the quasiparticle distribution near the Fermi surface. There is an additional radial degree of freedom that parametrizes different decay modes within an angular sector, but only a single mode per sector becomes long-lived with remaining modes having much shorter Fermi-liquid lifetime. Additional work has focused on the empirical signatures of tomographic transport~\cite{ledwith19b,hofmann_collective_2022,hofmann24,nazaryan24,kryhin25,rostami25,benshachar25a,benshachar25b,estrada25,starkov25,maki25}. A particular focus is on anomalous scaling signatures in temperature or wave number~\cite{ledwith19b,kryhin25,rostami25}, although this can be difficult to experimentally confirm with certainty since they only emerge at asymptotically low temperatures (i.e., when a plethora of long-lived modes contribute) where electron-interaction dominated transport tends to be suppressed by impurity scattering. More recent work has aimed to establish signatures of a general separation in the lifetimes of even and odd modes without requiring a particular parametric scaling~\cite{benshachar25a,benshachar25b}. In particular, as a hallmark signature of tomographic transport, it has been predicted that any tomographic  effect should be quickly suppressed even by weak magnetic fields~\cite{rostami25,benshachar25a,benshachar25b}. In exciting recent experimental developments, two works purport to find evidence of tomographic transport in the lifetime of cyclotron resonances~\cite{moisenko25} and the temperature scaling of the electron viscosity~\cite{zeng24}. Quite generally, the advancement of probes that locally image the flow of charge~\cite{sulpizio19,vool21,aharonsteinberg22,krebs_imaging_2023} is likely to result in the unambiguous identification of tomographic transport in the near future.

However, a basic and seemingly obvious question remains: Is the tomographic even-odd effect merely an anomaly specific to two-dimensional Fermi liquids or is it a more general feature? In particular, does it exist for three-dimensional Fermi liquids? In spite of the fact that solutions of the electron collision integral for 3D Fermi liquids in the context of specific transport lifetimes date back to the seminal work by~\textcite{abrikosov59}, and~\textcite{brooker_transport_1968}, and \textcite{sykes_transport_1970}, and are standard textbook material~\cite{nozieres_theory_1999, baym91,smith89}, a systematic description of different quasiparticle decay modes that captures the even-odd effect has not been given. Recent theoretical work~\cite{liao_two-dimensional_2020, buterakos_presence_2021, das_sarma_know_2021, ahn_fragile_2021} has extended the theories of single-quasiparticle decay rate and effective mass renormalization to higher orders in both energy and temperature in 2D and 3D Fermi liquids, discovering extremely complex coupled nonanalytic logarithmic terms in the higher-order self-energy of interacting Fermi liquids. However, this involved single-particle many-body effects, quite distinct from the tomographic dynamics of interest in the current work.
In this paper, we show that 3D Fermi liquids do indeed exhibit the even-odd effect, in contrast to conventional wisdom.

To frame the following discussion, let us review the conventional argument that leads to the absence of odd-even staggering of relaxation rates in 3D Fermi liquids, which relies on the fact that head-on scattering is not enforced by Pauli exclusion in 3D, distinct from the 2D case. This is best understood from a geometry argument. For an isotropic Fermi liquid in any dimension or at any temperature (degenerate or non-degenerate), energy and momentum conservation imply that the momentum transfers \mbox{$\bf{k} = \bf{p}_1-\bf{p}_1' = \bf{p}_2'-\bf{p}_2$} and ${\bf p}_1 - {\bf p}_2'$ between the initial momentum and the two final-state vectors are perpendicular:
\begin{align}\label{eqn:angle_fixed}
    {\bf k} \cdot ({\bf p}_1 - {\bf p}_2') &= 0 .
\end{align}
At low temperatures, where all momenta are close to the Fermi surface, in 2D this is enough to fix either (1) a forward scattering process, \mbox{$\bf{k}=0$}, (2) an exchange process, \mbox{$\bf{p}_1-\bf{p}_2'=0$}, or (3) head-on scattering, \mbox{$\bf{p}_2 = -\bf{p}_1$} and \mbox{$\bf{p}_2'=-\bf{p}_1'$}. The first two processes do not relax the quasiparticle distribution, leaving only head-on scattering as a relaxation channel. This is sketched in Fig.~\ref{fig:2d_vs_3d}(a), where the scattering momenta mark the corners of a rectangle. By the argument sketched above, head-on scattering then induces a parity-dependence in relaxation rates. In contrast, in the 3D case that we illustrate in Fig.~\ref{fig:2d_vs_3d}(b), there is an additional free parameter $\varphi_2$ that describes the change in the azimuthal angle between the scattering momenta, and head-on scattering will only occur in the special case \mbox{$\varphi_2=\pi$}. For this reason, a parity-odd deformation has a priori no obstruction to being relaxed at the usual \mbox{$\sim T^2/\hbar T_F$} Fermi-liquid scattering rate.
Indeed, we confirm this scaling in our calculations such that there is thus no parametric in temperature separation between the even and odd-parity scattering rates in 3D.

However, there is a gap in the above argument: It does not rule out an even-odd effect in the prefactor of these $T^2/\hbar T_F$ rates. This is precisely what we find for a wide range of electron-electron interactions, with the amplitude of  odd-mode scattering rates being suppressed relative to those of even-mode scattering rates. In particular, for the reasonable assumption of an electron-electron interaction that is approximately constant in momentum, e.g., because of strong screening, we find that the lowest-order odd-parity scattering rate is only $60\%$ as large as the lowest-order even-parity scattering rate, indicating that this 3D even-odd effect is quite large. There are in fact several additional considerations that could further enhance the effect in 3D: Since Fermi temperatures in 3D metals are on the order of $10^4$K, much higher than those in typical 2D systems that are on the order of \mbox{$1{\rm -}100$K}, there will be a large absolute difference between the even and odd rates for a fixed fraction of the Fermi temperature. Likewise, the accessible temperature range over which the even-odd separation exists will be generically larger in 3D. Additionally, we find that interactions that prefer large-angle scattering can further enhance the separation between even-odd scattering rates. While such interactions might not be common, the wide variety of hydrodynamic material candidates makes their realization plausible because of the background lattice and band structure effects in specific materials.

Having established that, contrary to the prior consensus, 3D Fermi liquids can also manifest collective tomographic behavior arising from electron-electron interactions, we then turn to the question of experimental probes of this effect. Here, it is necessary to consider correlators beyond the hydrodynamic long-wavelength limit in order to generate a microscopic response that includes higher-order odd-parity modes~\cite{ledwith19b,hofmann_collective_2022,nilsson_nonequilibrium_2024}. We demonstrate that just as in 2D the transverse conductivity $\sigma_\perp(\omega,q)$ contains signatures of the 3D even-odd effect, with a tomographic parameter regime in the static transverse conductivity at finite wavelengths. Moreover, transverse collective modes, just as in 2D, have relaxation rates set by the relaxation rate of the odd-parity modes. Studying the transverse conductivity in this limit thus provides a route to the identification of the slow odd-mode scattering in~3D. 

The paper is organized as follows: In Sec.~\ref{sec:methods_results}, we give an overview of our results, discussing both how to determine the scattering rates in an angular basis for a generic electron-electron interaction and reviewing transport signatures of the even-odd effect in 3D. In Sec.~\ref{sec:coll_int_three}, we give more details on the diagonalization of the collision integral in three dimensions. In Sec.~\ref{sec:exp_sign}, we discuss using both the static transverse conductivity and the collective modes of the transverse conductivity as a probe of the even-odd effect in 3D. Finally, in Sec.~\ref{sec:conc_outlook}, we review our conclusions and give an outlook for future results.

\section{Results}
\label{sec:methods_results} 

\subsection{Quasiparticle decay rates in three-dimensional Fermi liquids}

The main results of this paper are the decay rates~$\gamma$ for a three-dimensional isotropic Fermi liquid, which set the relaxation of small deviations $\delta f$ in the quasiparticle distribution from equilibrium $f_0(\bf{p})$. We parametrize these deviations 
in terms of a small momentum-dependent perturbation $T\psi$ to the local chemical potential,
\begin{equation} \label{eqn:deltaf_psi}
\delta f(t,\bf{r},\bf{p}) = \left(-\frac{\partial f_0}{\partial (\beta \varepsilon_{\bf{p}})}\right)\psi(t,\bf{r},\bf{p}).
\end{equation}
The prefactor $(-\partial f_0/\partial \varepsilon)$, which is sharply peaked at the Fermi surface, restricts the perturbations to the vicinity of the Fermi surface. For a rotationally symmetric Fermi surface, by the Wigner-Eckart theorem~\cite{arfken_mathematical_2013},  these perturbations can be classified in a basis of spherical harmonics~$Y_{lm}$. We thus separate the angular dependence of the momentum (in terms of the polar angles $\theta$ and $\varphi$) and the radial dependence (in terms of the magnitude~$|{\bf p}|$)  and decompose
\begin{equation}\label{eqn:sph_harm_basis}
 \psi(\bf{p}) = \sum_{n,l=0}^{\infty}\sum_{m=-l}^{l}\psi_{nlm}u_n(p)Y_{lm}(\theta,\varphi) ,
\end{equation}
where $Y_{lm}(\theta,\varphi)$ are spherical harmonics that depend on integer parameters $l\geq 0$ and \mbox{$|m|\leq l$}, and $u_n(p)$ with a radial index $n$ parametrize the dependence on the magnitude of the momentum~$p$. We review standard conventions and identities for~$Y_{lm}$ in Appendix~\ref{app:identities}. The expansion~\eqref{eqn:sph_harm_basis} is similar to 2D, where perturbations are decomposed in a basis of angular modes $e^{im\theta}$ with an angular index~$m$. As for 2D~\cite{hofmann_collective_2022,nilsson_nonequilibrium_2024}, the leading-order temperature scaling~\mbox{$\sim T^2/(\hbar T_F)$} of the longest-lived mode is obtained by setting the radial function~$u_n$ to a constant, i.e., with a perturbation that describes an angle-dependent rigid shift of the Fermi surface. Our aim is thus to determine the decay rates $\gamma_l$ of different angular modes, which, again by the Wigner-Eckart theorem, will also be independent of~$m$. Note that in 2D parity-even (odd) perturbations are those with even (odd) $m$, while in 3D the parity-even (odd) perturbations are those with even (odd) $l$.

\begin{figure}
    \centering
    \includegraphics[width=\columnwidth]{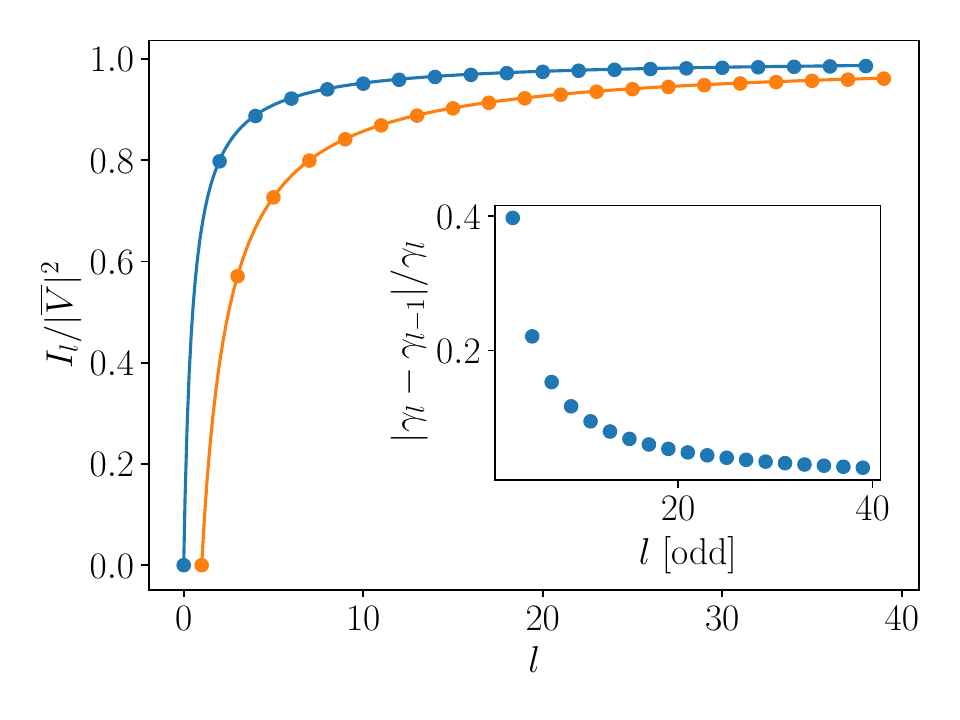}
    \caption{Decay rate coefficient \mbox{$I_l$} of the longest-lived modes in a 3D Fermi liquid with constant interaction potential as a function of the angular index~$l$, Eq.~\eqref{eqn:constant}. Blue points indicate decay rates for even $l$ and orange points for odd $l$. The smooth curves are drawn to guide the eye. The inset shows the relative difference \mbox{$|\gamma_l-\gamma_{l-1}|/\gamma_l$} of odd and even relaxation rates for odd $l\geq 3$. The difference peaks at $40\%$ for $l=3$ and decreases at large $l$.}
    \label{fig:const_pot}
\end{figure}

\begin{figure*}
    \centering
    \includegraphics[width=0.45\textwidth]{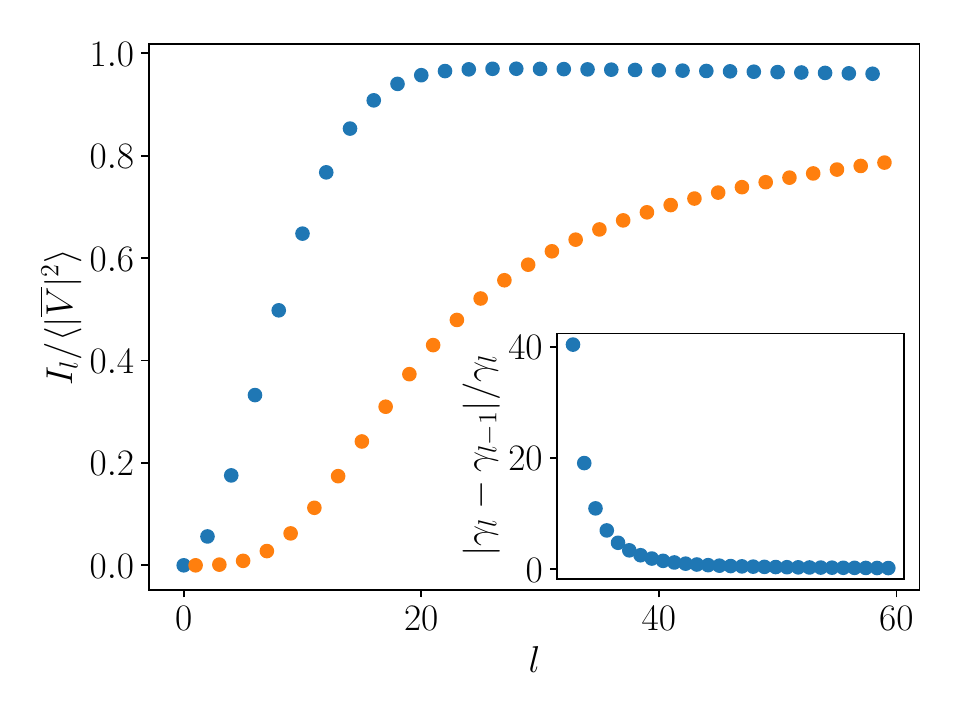}
    \includegraphics[width=0.45\textwidth]{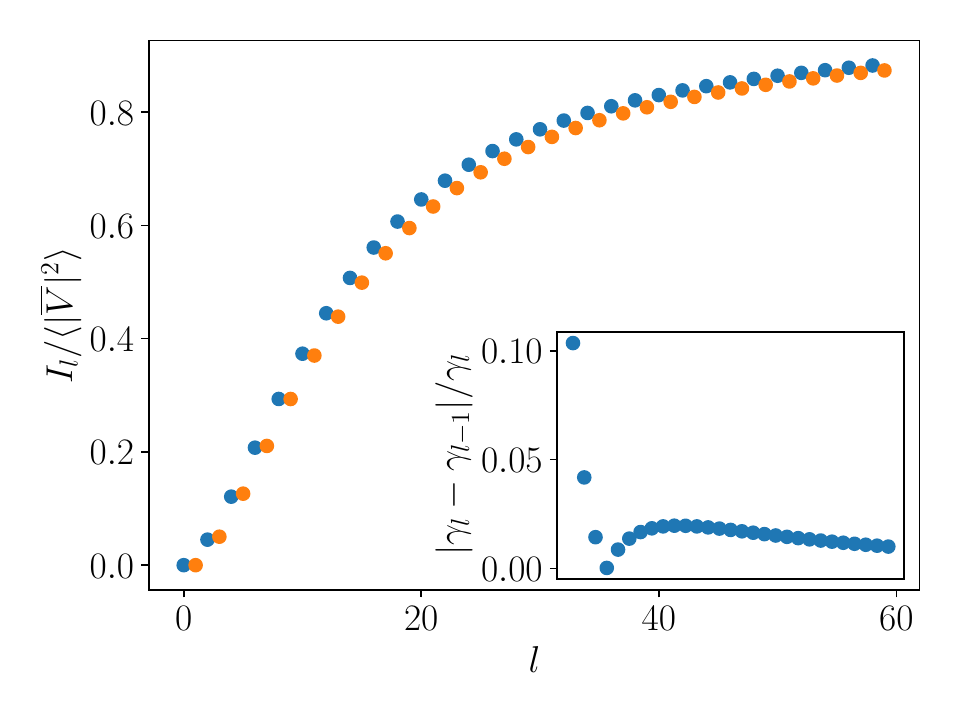}
    \caption{Decay rate coefficient \mbox{$I_l$} of the longest-lived modes in a 3D Fermi liquid for an interaction that prefers (a)~large-angle scattering [Eq.~\eqref{eqn:large_angle} with \mbox{$\sigma = 0.01$}] and (b)~small-angle scattering [Eq.~\eqref{eqn:defn_screened_coulomb} with~\mbox{$r_s = 0.01$}], where \mbox{$\langle|\overline{V}|^2\rangle$} is defined in Eq.~\eqref{eqn:defn_ovVsq}. Even (odd) mode scattering rates are shown in blue (orange). The insets display the relative difference between even and odd mode scattering for odd \mbox{$l\geq 3$}. (a) Large-angle scattering enhances the difference between even and odd rates. The inset shows a relative difference between odd and even rates as large as~$40$. (b) Small-angle scattering suppresses the difference between even and odd scattering rates, but even here a  relative difference of $10\%$ remains.}
	\label{fig:large_small}
\end{figure*}

To leading order in temperature, we find that
\begin{equation} \label{eqn:gamma_scale}
\gamma_l = \frac{\pi}{6}\frac{T^2}{\hbar T_F}N_fI_l,
\end{equation}
where $N_f$ is the number of flavors, including spin and valley degeneracy, and $I_l$ is a numerical constant. As anticipated in the introduction, the decay rate will always scale like $T^2$ at low temperatures. For a constant interaction, the coefficient $I_l$ can be calculated exactly and is given by
\begin{equation} \label{eqn:constant}
I_l = 2 \frac{N_f-1}{N_f} \, \overline{V}^2 \begin{cases} \dfrac{2l}{2l+1} &\mbox{if } l \ {\rm even}, \\[2ex] \dfrac{2l-2}{2l+1} &\mbox{if } l \ {\rm odd},\end{cases}
\end{equation}
where $\overline{V}^2$ is a dimensionless constant giving the strength of interelectron repulsion. Note that we will often take \mbox{$N_f=2$} to account for spin degeneracy unless otherwise mentioned. 
This is plotted in Fig.~\ref{fig:const_pot}, with details of the derivation in Sec.~\ref{sec:coll_int_three}.  
As is apparent from the figure, the decay rates display an even-odd effect at small~$l$. In particular,~$\gamma_3$ is only around $70\%$ as large as~$\gamma_2$ and around $60\%$ as large as~$\gamma_l$ for large values of~$l$. As discussed in the introduction, the appearance of an even-odd effect in 3D is surprising since there is no phase space restriction that enforces head-on scattering at low temperature. As a check of our result, we note that \mbox{$I_0=I_1=0$}. Physically, this is a consequence of the fact that $l=0$ represents a change in the Fermi surface volume, which cannot decay because of mass conservation, while $l=1$ represents a translation of the Fermi surface, which cannot decay because of momentum conservation.

In more detail, to investigate the effect of a generic nonconstant interaction potential $V(p)$, we obtain the following simple integral expression:
\begin{align}
I_l =& \  8\int_0^{1}{\rm d}(\cos \theta_1)\int_0^{2\pi}\frac{{\rm d}\varphi_2}{2\pi}\ |\overline{V}|^2 \nonumber \\
&\times \Biggl[\sum_{\substack{m=1 \\ l-m \ {\rm odd}}}^{l-1} \frac{(l-m)!}{(l+m)!}P^{m}_l(\cos\theta_1)^2\sin^2\left(\frac{m\varphi_2}{2}\right)\Biggr],   \label{eqn:defn_Il}
\end{align}
where $|\overline{V}|^2$ is a dimensionless scattering matrix element given by
\begin{align} \label{eqn:defn_ovVsq}
|\overline{V}|^2 =& \  \overline{V}^2(\cos\theta_1) + \overline{V}^2\left(\sin\theta_1\sin[\varphi_2/2]\right) \nonumber \\[0.5ex]
&-\frac{2}{N_f}\overline{V}(\cos\theta_1)\overline{V}\left(\sin\theta_1\sin[\varphi_2/2]\right) 
\end{align}
with
\begin{equation} \label{eqn:defn_dimV}
\overline{V}(x) = \frac{m^*p_F}{2\pi \hbar^2}V(2p_Fx) .
\end{equation}
The angles are as shown in Fig.~\ref{fig:2d_vs_3d}(b), where the interaction matrix element parametrizes the momentum transfer between the initial and the final states, with~\mbox{$|{\bf k}| = 2 p_F \cos \theta_1$} and~\mbox{$|\bf{p}_1'-\bf{p}_2| = 2p_F\sin\theta_1|\sin\varphi_2/2|$} [cf.~Eq.~\eqref{eqn:angle_fixed}]. Using a constant interaction in Eq.~\eqref{eqn:defn_Il} gives Eq.~\eqref{eqn:constant}. It can be checked for all interactions that~\mbox{$I_0 = I_1 = 0$}, just as for the case of a constant interaction. Furthermore, at large~$l$, the coefficient is on the order of the angle average
\begin{align}
    \langle |V^2| \rangle = \int_0^{1}{\rm d}(\cos \theta_1)\int_0^{2\pi}\frac{{\rm d}\varphi_2}{2\pi}\ |\overline{V}|^2 .
\end{align}

To make the connection to the 2D calculation, we note that if $|\overline{V}|^2$ were an interaction that allowed only for head-on scattering, \mbox{$\varphi_2=\pi$}, then Eq.~\eqref{eqn:defn_Il} reveals that \mbox{$I_l = 0$} for odd $l$. Such an exact cancellation of the odd-mode scattering rates at order $T^2$ would then imply a parametric-in-temperature separation between even and odd-parity scattering, just as in two dimensions. We note that it might be challenging to engineer an interaction that prefers \mbox{$\varphi_2=\pi$} without affecting $\theta_1$ due to their combined appearance in Eq.~\eqref{eqn:defn_ovVsq}. Nonetheless, we find that an interaction that prefers large-angle scattering also prefers \mbox{$\varphi_2\sim \pi$}, which (as we confirm below) enhances the separation between even and odd scattering. 

\begin{figure*}
    \centering
    \includegraphics[width=0.45\textwidth]{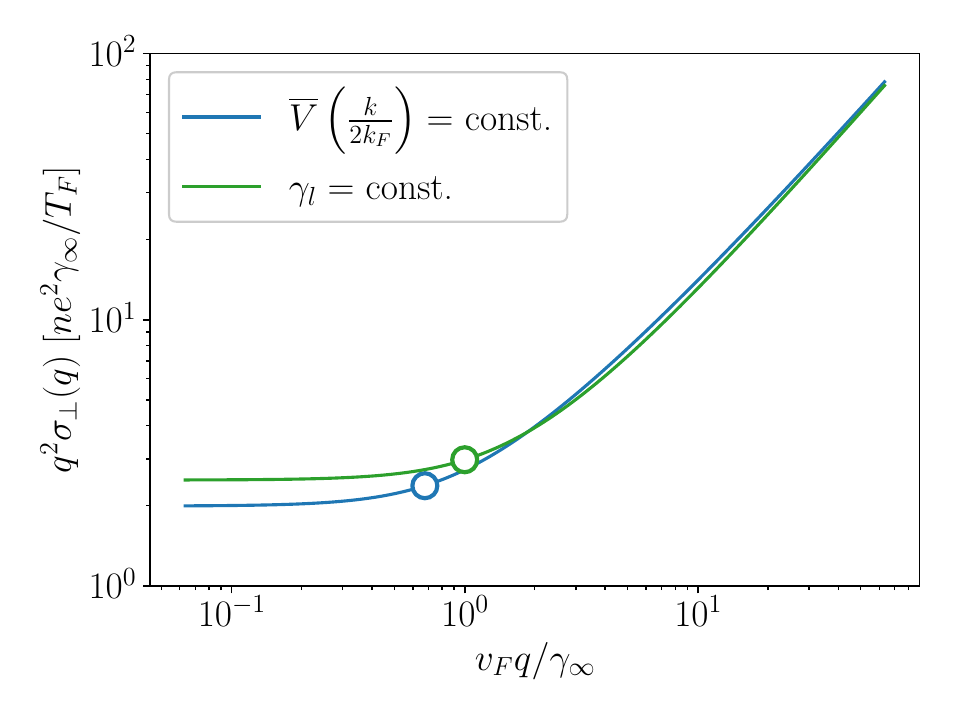}
    \includegraphics[width=0.45\textwidth]{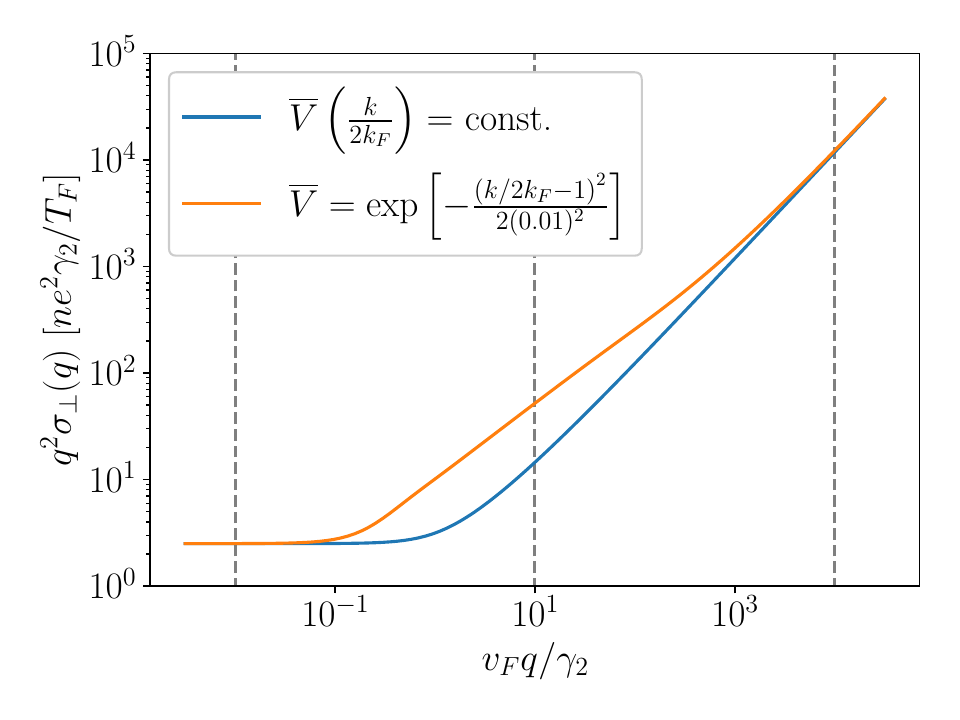}
    \caption{Scaled static transverse conductivity $q^2 \sigma_\perp(q)$ as a function of the wave vector $q$. (a) Static conductivity for a constant interaction potential (blue curve), which displays the even-odd effect [cf.~Eq.~\eqref{eqn:constant}], and a constant scattering rate $\gamma_l = \gamma$ for $l \geq 2$ (green curve). The wave numbers where corrections to hydrodynamic scaling appear are marked with a circle. For an even-odd staggering of decay rates, finite-wavelength corrections appear at significantly smaller wave numbers. (b) Static conductivity for with constant interactions [Eq.~\eqref{eqn:constant}] (blue curve), and large-angle interactions [Eq.~\eqref{eqn:large_angle}] (orange curve). The corresponding scattering rates are shown in Fig.~\ref{fig:large_small}(a). Gray dashed lines indicate the values of $q$ for which the microscopic deformations of the Fermi surface are plotted in Fig.~\ref{fig:static_deformations}. Note that compared to panel (a), the axes are rescaled by $\gamma_2$ so the different scaling behavior of the large-angle scattering can be cleanly observed.}
	\label{fig:static_cond}
\end{figure*}

We demonstrate this additional interaction dependence on the even-odd staggering of relaxation rates in Fig.~\ref{fig:large_small}. First, Fig.~\ref{fig:large_small}(a) shows the decay rates for an interaction potential that favors large-angle scattering, which in dimensionless form is given by
\begin{align}\label{eqn:large_angle}
    \overline{V}_{\rm la}(x) &= V_0 e^{-(x-1)^2/(2\sigma^2)} .
\end{align}
In real space, this model potential scales as~\mbox{$V(r) = DV_0 \sin(2p_Fr)/(2p_Fr)$} for~\mbox{$r\ll (2p_F\sigma)^{-1}$} and as \mbox{$V(r) = DV_0 \sigma^3\sqrt{\pi/2}\exp[-(2p_Fr\sigma)^2/2]$} for \mbox{$r\gg (2p_F\sigma)^{-1}$}, where \mbox{$D = 16\varepsilon_F/\pi$}. Figure~\ref{fig:large_small}(a) shows the decay rates with~\mbox{$\sigma=0.01$}, with a very clear enhancement of the parity effect in the relaxation rates, with even-parity rates exceeding the odd-parity rates by a factor of $40$ for the chosen parameter value. 

Likewise, in Fig.~\ref{fig:large_small}(b), we show the decay rates for a screened Coulomb interaction with \mbox{$V(p) = e^2/\varepsilon_0 (p^2+p_{TF}^2)$}, where \mbox{$p^2_{TF} = e^2m^*p_F/(\varepsilon_0 \pi^2\hbar^2)$} is the Thomas-Fermi wave vector. In nondimensionalized form
\begin{equation}\label{eqn:defn_screened_coulomb}
\overline{V}_{\rm Cou}(x) = \frac{\pi}{2}\frac{p^2_{TF}}{(2p_Fx)^2 + p_{TF}^2}.
\end{equation}
Here the dimensionless factor \mbox{$p_{TF}^2/4p_F^2 = 0.17r_s$}, where~$r_s$ is the Wigner-Seitz radius divided by the Bohr radius, proportional to the ratio of the Coulomb repulsion to kinetic energy~\cite{girvin2019modern}. For large interaction parameters \mbox{$r_s\gg 1$}, $|\overline{V}|^2$ is approximately constant over the range of integration and we obtain Eq.~\eqref{eqn:constant}. For small interactions $r_s\ll 1$, in contrast, $|\overline{V}|^2$ will enhance small-angle scattering. 
Based on the arguments above, we might expect this to reduce the even-odd effect in three dimensions. Indeed, this is what we find in Fig.~\ref{fig:large_small}(b), where we show the decay rates with a screened Coulomb interaction for a very small~\mbox{$r_s=0.01$}. We note that even for this small parameter value, the even-odd effect is still as large as~$10\%$. Screened Coulomb interactions in typical materials have \mbox{$r_s \simeq 1$}~\cite{girvin2019modern}, and we thus expect a large even-odd effect as described by Eq.~\eqref{eqn:constant}. 

\subsection{Experimental signature}

Having established the existence of an even-odd effect in 3D Fermi liquids, we proceed to show how such an effect might be detected. We specifically focus on the transverse conductivity, which is known to display signatures of odd-parity scattering rates in~2D~\cite{ledwith19b, hofmann_collective_2022, rostami25, kryhin25}.

\begin{figure*}
    \centering
    \includegraphics[width=0.9\textwidth]{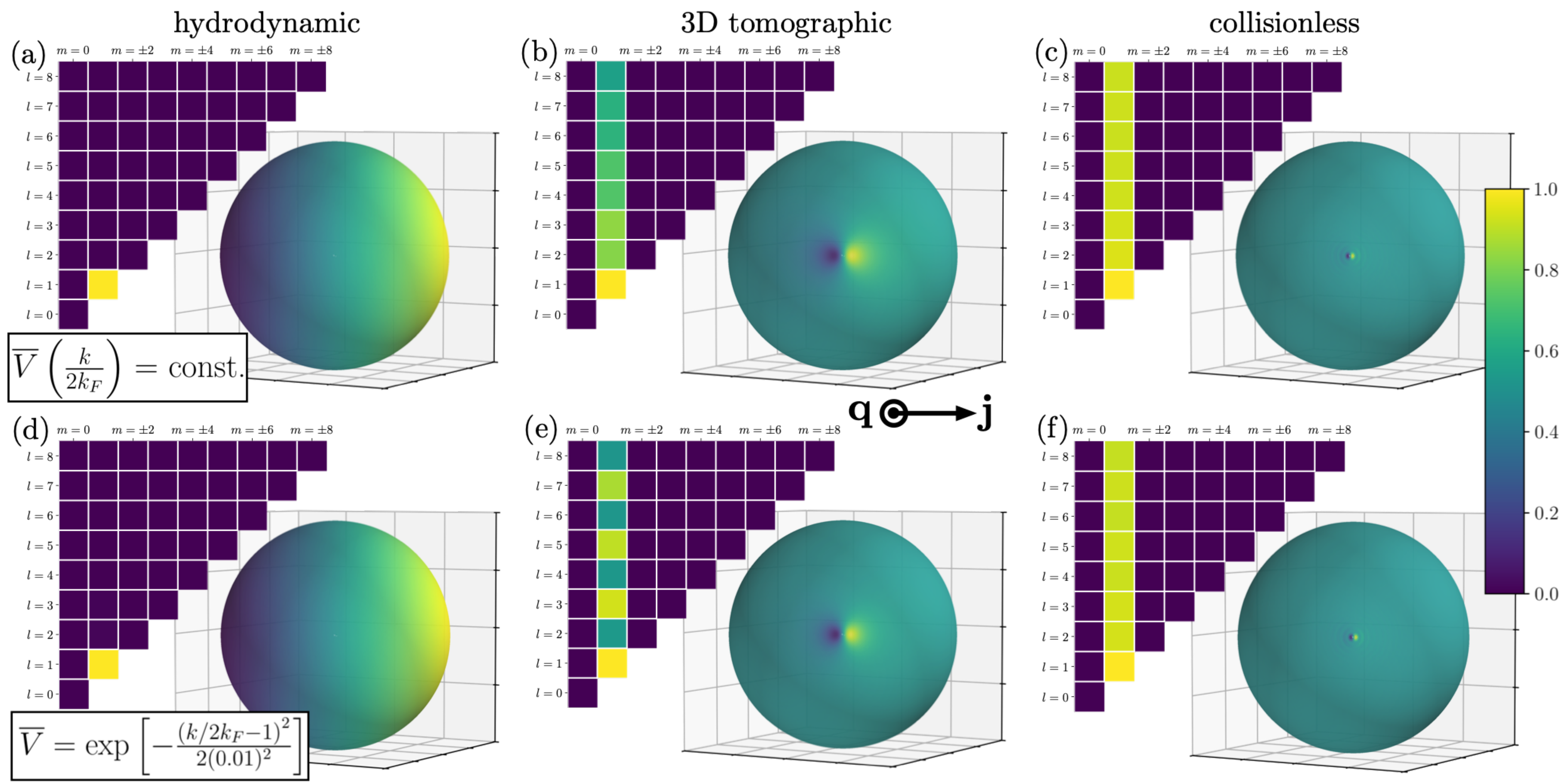}
    \caption{Fermi surface deformations plotted on the Fermi surface, with squares displaying the decomposition into spherical harmonics. The current $\bf{j}$ points to the right, while the wave vector $\bf{q}$ points out of the page. The first row assumes a constant interaction [Eq.~\eqref{eqn:constant}], and the second row assumes a large-angle interaction [Eq.~\eqref{eqn:large_angle}]. The columns correspond, from left to right, to the wave number indicated by gray dashed lines in Fig.~\ref{fig:static_cond}. The hydrodynamic limit $v_Fq\ll \gamma_l$ is the first column, the 3D tomographic limit is the second column, and the collisionless limit $\gamma_l \ll v_Fq$ is the third column. In the hydrodynamic limit, only the conserved current is excited, while in the collisionless limit all modes are excited with a deformation that is strongly peaked about $\bf{q}$. In the tomographic like limit, the odd modes exhibit enhanced occupation, especially clear for the large-angle scattering in panel~(e).}
	\label{fig:static_deformations}
\end{figure*}

We begin with the static transverse conductivity $\sigma_\perp(q)$, which we obtain from a numerical solution of the Boltzmann equation at zero frequency [see Sec.~\ref{sec:exp_sign} for details]. This could be probed experimentally either directly, by subjecting the system to a spatially varying electric field, or indirectly by measuring flow through a constriction or a channel. In the latter case, just as in 2D~\cite{moll16,ledwith19b,nazaryan24,rostami25,kryhin25}, the width of the constriction~$W$ would set a tunable momentum scale $\sim 1/W$ to allow for probing of the static conductivity. Care must be taken, however, that this level of approximation misses effects from boundary layers and slip boundary conditions~\cite{benshachar24a,benshachar24b,benshachar25a,benshachar25b}.

Our results for the static conductivity are plotted in Fig.~\ref{fig:static_cond}. For all choices of scattering rates, we find that when $v_Fq$ is smaller than all other scattering rates (referred to as the hydrodynamic limit),
\begin{equation}
\lim_{q\rightarrow 0} q^2\sigma_\perp(q) = \frac{5}{2}\frac{ne^2\gamma_2}{T_F}.
\end{equation}
The expectation is that corrections to this constant scaling appear when $v_Fq$ is on the order of $\sqrt{\gamma_2\gamma_3}$. This is apparent in Fig.~\ref{fig:static_cond}(a), where it can be seen that corrections appear for values of $v_Fq$ that are $30\%$ smaller for scattering rates given by Eq.~\eqref{eqn:constant} compared to an ansatz with constant scattering rates.
In the opposite limit, when $v_Fq$ is larger than all other scattering rates (i.e., the collisionless limit), $q^2\sigma_\perp(q)$ scales linearly with $q$ as 
\begin{equation}
\lim_{q\rightarrow \infty} q^2\sigma_\perp(q) \rightarrow 1.18\frac{ne^2 v_Fq}{T_F}.
\end{equation}
In the intermediate regime where $v_Fq$ is larger than some of the odd-mode scattering rates but smaller than the even mode scattering rates,  $q^2\sigma_\perp(q)$ is neither a constant nor scales linearly with $q$. Instead, it exhibits a crossover behavior, which is displayed in Fig.~\ref{fig:static_cond}(b). In two dimensions, this intermediate regime is known as the tomographic regime.
Our plot of the static conductivity also shows an intermediate regime with scaling of $q^2\sigma_\perp(q)$ as a function of $q$ in between the constant hydrodynamic and linear collisionless scaling. This effect can be seen to be most pronounced for the interaction given by Eq.~\eqref{eqn:large_angle}, since the preference for large-angle scattering leads to the highly suppressed odd-mode scattering rates seen in Fig.~\ref{fig:large_small}. While the precise scaling exponent will be different for different interactions, the presence of an intermediate scaling regime in the static conductivity offers a signature of the even-odd effect in 3D; in this regime, the odd mode deformations can be expected to be enhanced.

To understand the physics of these limits better, we show in Fig.~\ref{fig:static_deformations} the microscopic Fermi surface deformations induced by the electric field in each limit. 
We see for both interactions that in the hydrodynamic limit [Figs.~\ref{fig:static_deformations}(a) and \ref{fig:static_deformations}(d)], only the conserved current is excited (i.e., \mbox{$l=1$}), as expected. In contrast, in the intermediate regime where $v_Fq$ is larger than some of the odd-mode scattering rates but smaller than the even-mode scattering rates, the deformations for both interactions display enhanced excitation of the odd modes compared to the even modes [Figs.~\ref{fig:static_deformations}(b) and \ref{fig:static_deformations}(e)]. This effect is most pronounced for the interaction given by Eq.~\eqref{eqn:large_angle}, since the preference for large-angle scattering leads to the highly suppressed odd-mode scattering rates seen in Fig.~\ref{fig:large_small}. Finally, in the collisionless limit, all modes are excited [Figs.~\ref{fig:static_deformations}(c) and \ref{fig:static_deformations}(f)].

Another signature of the odd-mode scattering rates in 2D lies in the collective modes of the transverse conductivity \cite{hofmann_collective_2022}. We follow the approach of Ref.~\cite{hofmann_collective_2022} and investigate these analytically. In order to extract the essential physics without committing to a particular form of the interelectron interaction $V$, we take the simplifying assumption that the decay rates do not scale with~$l$ beyond their even-odd dependence, i.e., that $\gamma_l =0$ for \mbox{$l=0$}, $\gamma_i$ for \mbox{$l=1$}, $\gamma$ for \mbox{$l\geq 2$} and even, and $\gamma'$ for \mbox{$l\geq 3$} and odd.
Here, the $l=1$ mode can only relax through momentum nonconserving processes such as Umklapp or impurity scattering set by $\gamma_i$. We assume that \mbox{$\gamma_i < \gamma' < \gamma$}, i.e. that $\gamma_i$ is the smallest scale and odd-parity modes are scattered at a lower rate than even-parity modes. With this ansatz for scattering rates, we conclude that, just as in 2D, the collective modes of the transverse conductivity can be expected to display the even-odd effect in 3D. In particular, in the limit \mbox{$\gamma_i \ll \gamma'\ll v_Fq\ll \gamma$} and \mbox{$\gamma \gamma' \ll (v_Fq)^2$}, the transverse conductivity exhibits a collective mode with damping \mbox{$\omega = -i\gamma'$}, i.e., governed by the odd-parity scattering rate. In contrast, in the limit \mbox{$\gamma_i \ll v_Fq\ll \gamma' \ll \gamma$}, higher-order odd-parity modes are not excited, leaving only the conserved current zero mode. Thus, we suggest that in the limit \mbox{$\gamma_i \ll \gamma'\ll v_Fq\ll \gamma$} and \mbox{$\gamma \gamma' \ll (v_Fq)^2$} the transverse conductivity will exhibit a collective mode whose frequency and Fermi surface deformation allow for identification of the odd-parity scattering rates in 3D Fermi liquids.

\section{Decay rates in angular basis}
\label{sec:coll_int_three} 

In this section, we summarize details on the evaluation of the linearized collision integral and the derivation of the Fermi liquid decay rates stated in Eqs.~\eqref{eqn:constant} and~\eqref{eqn:defn_Il}. 

\subsection{Linearized collision integral}

Microscopically, quasiparticle relaxation is described by the collision integral~${\cal J}$. In the dilute limit that is applicable to electron gases, this collision integral is set by two-particle scattering processes and  given by
\begin{align}\label{eqn:collision_int_initial}
&{\cal J}[f(t,\bf{p}_1)] = \ - \iiint \frac{d(\bf{p}_2,\bf{p}_1',\bf{p}_2')}{(2\pi)^9}
W({\bf p}_1', {\bf p}_2'|{\bf p}_1,{\bf p}_2) \nonumber\\
&\qquad\times \bigl[f({\bf p}_1)f({\bf p}_2)(1-f({\bf p}_1'))(1-f({\bf p}_2'))\nonumber \\[1ex]
&\qquad- f({\bf p}_1')f({\bf p}_2')(1-f({\bf p}_1))(1-f({\bf p}_2))\bigr],
\end{align}
where $W({\bf p}_1', {\bf p}_2'|{\bf p}_1,{\bf p}_2)$ is the (screened Coulomb or other) scattering matrix element that enforces momentum and energy conservation. An application of Fermi's golden rule gives the explicit expression
\begin{align}\label{eqn:Coulomb_scatt_mat}
W({\bf p}_1', {\bf p}_2'|{\bf p}_1,{\bf p}_2) =& \  \frac{2\pi}{\hbar}|V|^2 (2\pi)^3 \delta^{(3)}({\bf p}_1 + {\bf p}_2 - {\bf p}_1' - {\bf p}_2')\nonumber \\
&\times \delta(\varepsilon_{{\bf p}_1} + \varepsilon_{{\bf p}_2} - \varepsilon_{{\bf p}_1'} - \varepsilon_{{\bf p}_2'}) ,
\end{align}
where we assume an isotropic electron dispersion 
\begin{equation}
 \varepsilon_{\bf{p}} = \frac{\hbar^2|\bf{p}|^2}{2m^*}
\end{equation}
with $m^*$ the effective mass. We assume a spin and valley-independent interaction matrix element~\cite{gottfriedyan03}
\begin{align}\label{eqn:defn_coulomb_int}
|V|^2 =& \ V^2(\bf{p}_1'-\bf{p}_1) + V^2(\bf{p}_1'-\bf{p}_2) \nonumber \\[1ex]
&- \frac{2}{N_f} V(\bf{p}_1'-\bf{p}_1)V(\bf{p}_1'-\bf{p}_2) .
\end{align}
Here, the first two terms describe the direct contribution and the last term describes the exchange contribution to scattering. Note that with an increasing number of flavors, the exchange contribution becomes suppressed. Throughout, we assume that the potential is isotropic, i.e., that \mbox{$V(\bf{p}) = V(p)$}.

It can be checked directly that for a Fermi-Dirac distribution
\begin{equation}
f_0({\bf p}) = \frac{1}{1 + \exp(\beta(\varepsilon_{{\bf p}} - \mu))} ,
\end{equation} 
the collision integral~\eqref{eqn:collision_int_initial} obeys detailed balance and vanishes. We describe perturbations around this solution using the deviation function~$\psi$ as defined in Eq.~\eqref{eqn:deltaf_psi}. This allows to define a linearized collision integral in terms of~$\psi$
\begin{equation}\label{eq:defL}
{\cal L}[\psi(\bf{p})] = \frac{-{\cal J}[\delta f(\bf{p})]}{f_0(\bf{p})(1-f_0(\bf{p}))}.
\end{equation}
The calculation of decay rates then corresponds to the eigenvalue problem for this operator, i.e., we want to determine deformations $\psi$ with~\mbox{${\cal L}[\psi] = \gamma \psi$}.

\subsection{Inner product and basis}

To determine the eigenvalues of ${\cal L}$, we introduce the inner product
\begin{equation}\label{eqn:defn_inner_prod}
\langle\tilde{\psi}|\psi\rangle = \frac{(2\pi\lambda_T)^2}{p_F}\int \frac{d\bf{p}}{(2\pi)^3} 
\tilde{\psi}^{*}({\bf p}) f_0({\bf p})(1-f_0({\bf p})) \psi({\bf p}),
\end{equation}
where \mbox{$\lambda_T = \sqrt{2\pi \hbar^2\beta/m^*}$} is the thermal wavelength. With this choice, we define matrix elements of the linearized collision integral as
\begin{align} \label{eqn:lin_coll_mat}
\langle\tilde{\psi}|{\cal L}|\psi\rangle =& \  \frac{(2\pi \lambda_T)^2}{4p_F} \iiiint \frac{d(\bf{p}_1,\bf{p}_2,\bf{p}_1',\bf{p}_2')}{(2\pi)^{12}}\nonumber \\[1ex]
&\times 
W({\bf p}_1', {\bf p}_2'|{\bf p}_1,{\bf p}_2)F_{121'2'} \nonumber \\[1ex]
&\times [\tilde{\psi}^{*}({\bf p}_1)+\tilde{\psi}^{*}({\bf p}_2) - \tilde{\psi}^{*}({\bf p}_1') -\tilde{\psi}^{*}({\bf p}_2')]\nonumber \\[1ex]
&\times [\psi({\bf p}_1)+\psi({\bf p}_2) - \psi({\bf p}_1') -\psi({\bf p}_2')],
\end{align}
where we abbreviate
\begin{equation}\label{eqn:fermi_func_prod}
F_{121'2'} = f_0(\bf{p}_1)f_0(\bf{p}_2)(1-f_0(\bf{p}_1'))(1-f_0(\bf{p}_2')).
\end{equation}
It is straightforward to check that $\langle\tilde{\psi}|{\cal L}|\psi\rangle$ is positive semidefinite and thus has real positive eigenvalues, as necessary for decay rates.

As discussed in Eq.~\eqref{eqn:sph_harm_basis}, the spherical symmetry of the problem implies that the collision integral is diagonal in a basis of spherical harmonics. We can use the inner product~\eqref{eqn:defn_inner_prod} to choose the remaining radial deformations~$u_n$ to form an orthonormal basis, which is discussed in more detail in Appendix~\ref{app:orth_basis_decomp}. In the following, as discussed, we consider only the leading-order contribution, which is given at low temperatures by \mbox{$u_0(p)=1$} and which represents a uniform shift in chemical potential.

We already established in the introduction that the scalar operator ${\cal L}$ is symmetric under a global rotation of all momenta, which by the Wigner-Eckart theorem~\cite{arfken_mathematical_2013} implies that it must be diagonal in~$l$ and~$m$ with diagonal elements independent of $m$. In other words,
\begin{align}
    \langle u_{n'}Y_{l',m'}|{\cal L}|u_nY_{lm}\rangle = \delta_{ll'}\delta_{mm'} ({\cal L}_l)_{n',n} ,
\end{align}
where we chose this notation to illustrate the independence of the linearized collision integral on $m$. We note in passing that this diagonal structure of the collision integral can also be confirmed in an explicit calculation, which is done in Appendix~\ref{app:int_out_rot}.

\subsection{Evaluating the scattering rates}

We want to determine the smallest eigenvalue $\gamma_l$ of~$({\cal L}_l)_{n,n'}$, which sets the longest lived decay mode of a given angular mode. To this end, we consider the leading order contribution $({\cal L}_l)_{0,0}$, which, by the Rayleigh-Ritz principle, gives a strict upper bound on the smallest relaxation time, 
\begin{align}
\gamma_l \lesssim \frac{\langle Y_{lm} | {\cal L} | Y_{lm} \rangle}{\langle Y_{lm} | Y_{lm} \rangle} = ({\cal L}_l)_{0,0} .
\end{align}
The agreement with the true eigenvalue is expected to be very close at low temperatures, where the quasiparticle distribution is predominantly composed of a rigid Fermi surface deformation. This is a common approximation that captures the dominant contribution to the quasiparticle deformation at low temperatures~\cite{hofmann_collective_2022}, and its validity can be investigated in subsequent work by numerically solving the full collision integral, as was recently done in 2D~\cite{nilsson_nonequilibrium_2024}.

We summarize the evaluation of the matrix element in the following, with some details relegated to Appendix~\ref{app:calc_gamml}. The first step is to transform to the momentum transfer coordinates illustrated in Fig.~\ref{fig:2d_vs_3d}. We then follow the explicit calculation of Appendix~\ref{app:int_out_rot} to integrate out a global rotation such that $\bf{k} = \bf{p}_1 - \bf{p}_1'$ is oriented along the $z$ axis and $\bf{p}_1$ lies in the $xz$ plane. This removes all angular dependence of $\bf{k}$ and the dependence of $\bf{p}_1$ on the azimuthal angle $\varphi$. This gives
\begin{align}
&({\cal L}_l)_{0,0} = \frac{\hbar \beta}{2 m^*p_F (2l+1)} \sum_{m=-l}^{l}\int \frac{{\rm d}k{\rm d}p_1{\rm d}(\cos\theta_1) {\rm d}^3\bf{p}_2}{(2\pi)^3}\nonumber \\
&\times k^2p_1^2|V|^2 F_{121'2'}\delta(\varepsilon_{{\bf p}_1} + \varepsilon_{{\bf p}_2} - \varepsilon_{{\bf p}_1'} - \varepsilon_{{\bf p}_2'}) \nonumber \\[1ex]
&\times \left|Y_{lm}(\theta_1,0) + Y_{lm}(\theta_2,\varphi_2) - Y_{lm}(\theta_1',0)-Y_{lm}(\theta_2',\varphi_2)\right|^2 .\label{eqn:first_Ll00}
\end{align}
Here, $\theta_1'$ is a function of $p_1,k,\theta_1$ since $\bf{p}_1'=\bf{p}_1-\bf{k}$, and likewise for $\theta_2'$.

We now split the energy delta function by replacing~$\varepsilon_{\bf{p}_1'}$ and~$\varepsilon_{\bf{p}_2'}$ by the energy transfer $\hbar \omega = \varepsilon_{\bf{p}_1}-\varepsilon_{\bf{p}_1'} = \varepsilon_{\bf{p}_2'}-\varepsilon_{\bf{p}_2}$. When we do this, we find that the Fermi factor $F_{121'2'}$ becomes a function of only $p_1,p_2$, and $\omega$. In particular, if \mbox{$w_i = \beta(\varepsilon_{\bf{p}_i}-\mu)$} and \mbox{$W=\beta\hbar \omega$}, then
\begin{align}
F_{121'2'} =& \  \frac{1}{4}\frac{1}{\cosh\left[\frac{w_1+w_2}{2}\right]+\cosh\left[\frac{w_1-w_2}{2}\right]}\nonumber \\
&\times \frac{1}{\cosh\left[\frac{w_1+w_2}{2}\right]+\cosh\left[\frac{w_1-w_2}{2}-W\right]} \label{eqn:F_in_terms_of_w}.
\end{align}
This factor ensures that $w_1,w_2,W$ are all of order $1$, and thus at low temperatures \mbox{$p_{1,2} \sim p_F + p_F{\it O}(T/T_F)$}. Expanding \mbox{$\hbar \omega = \varepsilon_{\bf{p}_1}-\varepsilon_{\bf{p}_1-\bf{k}} = \varepsilon_{\bf{p}_2+\bf{k}} - \varepsilon_{\bf{p}_2}$}, we see that
\begin{equation}\label{eq:deltamomentum}
\cos\theta_i = \sigma_i\frac{k}{2p_F} + {\it O}\left(\frac{T}{T_F}\right),
\end{equation}
where \mbox{$\sigma_1 = \sigma_2' = +1$} and \mbox{$\sigma_2=\sigma_1'=-1$}. Thus, at low temperatures, integrating out the energy delta function will fix $\theta_i$ [cf.~Fig.~\ref{fig:2d_vs_3d}].

Returning to the integral in Eq.~\eqref{eqn:first_Ll00}, neither $|V|^2$ nor the spherical harmonics will be a function of $p_1,p_2,\omega$. We can therefore integrate out these factors, which gives
\begin{align}\label{eqn:first_see_Tsq0}
&({\cal L}_l)_{0,0} = \  \frac{m^{*3}T^2}{48\pi\hbar^7p_F}\sum_{m=-l}^{l}\frac{(l-m)!}{(l+m)!}\int {\rm d} k \frac{{\rm d}\varphi_2}{2\pi}|V|^2 \nonumber\\
&\quad \times \sin^2\left(\frac{m\varphi_2}{2}\right)\left[P^m_l\left(\frac{k}{2p_F}\right) - P^m_l\left(-\frac{k}{2p_F}\right)\right]^2.
\end{align}
Here, we have replaced the spherical harmonics with their full expression in terms of associated Legendre polynomials and simplified using the low temperature expression for $\cos\theta_i$ [Eq.~\eqref{eq:deltamomentum}].

We first note that in three dimensions $|V|^2$ depends both on \mbox{$\cos\theta_1 = k/2p_F$} and on $\varphi_2$. As is apparent from Fig.~\ref{fig:2d_vs_3d}, $\varphi_2$ controls the extent to which scattering is head on, with perfect head-on scattering at \mbox{$\varphi_2=\pi$}. It enters into $|V|^2$ via $|\bf{p}_1'-\bf{p}_2|$. At low temperatures, \mbox{$|\bf{p}_1'-\bf{p}_2| = 2p_F\sin\theta_1|\sin\varphi_2/2|$}~\footnote{For head-on scattering, $\varphi_2 = \pi$, this reproduces what is known in 2D \cite{nilsson_nonequilibrium_2024}.}. Thus, a preference for large-angle scattering, i.e., for \mbox{$|\bf{p}_1'-\bf{p}_2| \sim 2p_F$}, will also prefer head-on scattering with \mbox{$\varphi_2\sim \pi$}.

Finally, the expression~\eqref{eqn:first_see_Tsq0} shows that the lowest-order matrix element will always scale as $T^2$ times some coefficient. If we nondimensionalize by defining the nondimensional \mbox{$\overline{V} = p_FV/\lambda_T^2T$}, we arrive at Eqs.~(\ref{eqn:gamma_scale})--(\ref{eqn:defn_dimV}), which are obtained using properties of the associated Legendre polynomials~\cite{arfken_mathematical_2013} to simplify the resulting expression for~$I_l$.

\section{Experimental signatures}
\label{sec:exp_sign}

Having diagonalized the collision integral, we now understand how perturbations to the Fermi-Dirac distribution will relax in a 3D Fermi liquid with no forcing or spatial dependence. Next, we turn our focus to inducing these perturbations in the experiment. One standard way to drive a Fermi liquid out of equilibrium involves applying an electric field. In order to understand how the 3D Fermi liquid will respond to this external forcing, we must solve the full Boltzmann equation. 

\subsection{Setting up the Boltzmann equation}

The full time evolution of the Fermi liquid quasiparticle distribution function $f(t,\bf{r},\bf{p})$ is dictated by the Boltzmann transport equation
\begin{equation} \label{eqn:Boltzmann_transport}
\left[\frac{\partial}{\partial t} +\bf{F} \cdot \frac{\partial}{\partial (\hbar\bf{p})} + \frac{\partial \tilde{\varepsilon}}{\partial (\hbar\bf{p})}\cdot \frac{\partial }{\partial \bf{r}}\right]f(t,\bf{r},\bf{p}) = {\cal J}[f].
\end{equation}
The left-hand side describes the evolution of the distribution function due to streaming, while the right-hand side describes the changes to $f$ as a result of collisions. Here, $\tilde{\varepsilon}$ is the local quasiparticle energy that takes into account mean-field interaction corrections~\cite{nozieres_theory_1999}. 
In the following, we assume an external forcing~\mbox{${\bf F}(t,{\bf r}) = -e {\bf E}(t,{\bf r})$} by a time- and position-dependent  electric field  $\bf{E} = {\bf E}_0 e^{-i(\omega t - \bf{q}\cdot \bf{r})}$, with choice of coordinates \mbox{$\bf{q} = q\hat{z}$} and $\bf{E}_0 = E_0(\sin\theta_E, 0, \cos\theta_E)$. For a weak perturbation, the deformation $\psi_{\bf{p}}$ will show the same spatial and time dependence as the external field. We obtain the following relation at leading order in the perturbation:
\begin{equation}\label{eq:boltzmann0}
\frac{\partial \psi_{\bf{p}}}{\partial t} + e\beta(\bf{v}_{\bf{p}} \cdot \bf{E})+ \bf{v}_{\bf{p}}\cdot \frac{\partial \overline{\psi}_{\bf{p}}}{\partial \bf{r}} = -{\cal L}[\overline{\psi}_{\bf{p}}].
\end{equation}
Here, \mbox{$\bf{v}_{\bf{p}} = \partial \varepsilon_{\bf{p}}/\partial (\hbar \bf{p})$} is the Fermi velocity, and the overbar $\overline{\psi}_{\bf{p}}$ denotes the deviation from local equilibrium. As reviewed in Appendix~\ref{app:identities}, in the spherical harmonic basis of Eq.~\eqref{eqn:sph_harm_basis} this is linked to the deviation from global equilibrium by
\begin{equation}\label{eqn:sph_decomp_Landau}
\overline{\psi}_{\bf{p}} = \sum_{n,l=0}^{\infty}\sum_{m=-l}^{l}\left(1 + \frac{F_l}{2l+1}\right)\psi_{nlm}u_n(p)Y_{lm}(\theta,\varphi),
\end{equation}
where $F_l$ are the Landau parameters.

As in the derivation of the collision integral, we neglect the $|{\bf p}|$ dependence on $\psi_{\bf{p}}$ and include only its angular dependence on the Fermi surface. 
Then, we can use our result from the previous section to reduce the linearized collision integral to its value for the $l$th spherical harmonic. Since we have already assumed that $\psi$ has no dependence on $p$, this is exactly $\gamma_l = ({\cal L}_l)_{0,0}$. Then, we have that
\begin{align}
&(-\omega - i\gamma_l)\psi_{lm} + \frac{v_Fq}{2}\left(B^-_{lm}\psi_{l-1,m} + B^+_{lm}\psi_{l+1,m}\right)\nonumber \\[1ex]
&= \  -i\sqrt{\frac{2\pi}{3}}e\beta E_0v_F\begin{cases} \sqrt{2}\cos\theta_E &\mbox{if } l=1,m=0\\ \mp \sin\theta_E &\mbox{if } l=1,m=\pm 1\\ 0&\mbox{ otherwise,}\end{cases} \label{eqn:Boltzmann_tower},
\end{align}
where
\begin{align}
B^-_{lm} = \sqrt{\frac{l^2 - m^2}{l^2 - 1/4}}b_{l-1}&, B^+_{lm} = \sqrt{\frac{(l+1)^2 - m^2}{(l+1)^2 - 1/4}}b_{l+1}\nonumber,\\
b_l =& \  1+\frac{F_l}{2l+1}.\label{eqn:defn_Blm}
\end{align}
The factors of $B^{\pm}_{lm}$ arise from the decomposition of $\cos\theta \, Y_{lm}(\theta,\varphi)$ into  spherical harmonics. These expressions are summarized in Appendix~\ref{app:identities}.

Before outlining the solution of Eq.~\eqref{eqn:Boltzmann_tower}, a few remarks are in order. First, the $\cos\theta$ factor is symmetric under a rotation about the $z$ axis and thus can only couple different values of $l$, but not different values of $m$.  Different $m$ sectors of the distribution are thus decoupled in the kinetic equation in the formulation of Eq.~\eqref{eqn:Boltzmann_tower}, and we can solve the equations separately for every single~$m$. Since the electric field only excites~\mbox{$m=0,\pm 1$} modes, this implies that we can ignore all  higher values of~$m$ when evaluating the current response and take \mbox{$\psi_{lm} = 0$} if \mbox{$|m|\geq 2$}. Second, we note that Eq.~\eqref{eqn:Boltzmann_tower} takes a very similar form to the two-dimensional kinetic equation projected onto angular harmonic deformations~\cite{hofmann_collective_2022}, with~$l$ now in place of~$m$. The $B^{\pm}_{lm}$ terms only appear in three dimensions, however. They are well behaved as $l\rightarrow \infty$, neither going to zero nor diverging, but instead approaching $b_{l\pm 1}$.

\subsection{Solving the infinite tower}

Let us now solve the tower of Eq.~\eqref{eqn:Boltzmann_tower}. We start with a strategy of first determining $\psi_{lm}$ for \mbox{$l\geq 2$} and working our way backwards. Defining
\begin{equation} \label{eqn:defn_xlm}
x_{lm}(\omega,q) = \frac{\psi_{lm}(\omega,q)}{\psi_{l-1,m}(\omega,q)},
\end{equation}
Eq.~\eqref{eqn:Boltzmann_tower} becomes
\begin{equation} \label{eqn:xlm}
\frac{B^-_{lm}}{x_{lm}}-2\left(\frac{\omega}{v_Fq} + i\frac{\gamma_l}{v_Fq}\right)+B^+_{lm}x_{l+1,m} = 0 , \ l\geq 2.
\end{equation}
This equation is solved via continued fractions~\cite{hofmann_collective_2022}. Inverting the expression gives for $l\geq 2$ 
\begin{equation}
x_{lm} = \frac{B^{-}_{lm}}{2\alpha_{l}-B^+_{lm}x_{l+1,m}}, \ \alpha_l = \frac{\omega}{v_Fq} + i\frac{\gamma_l}{v_Fq} . \label{eqn:cont_frac}
\end{equation}
This expression can be solved numerically via Lentz's algorithm \cite{hofmann_collective_2022}. For analytical estimates, it is helpful to bring the expression~\eqref{eqn:cont_frac} to a form similar to the 2D case by rescaling away  the $B^{\pm}_{lm}$ factors, which will shift all of the $l$ dependence into the $\alpha_l$ terms. 
To this end, define \mbox{$\tilde{x}_{lm} = x_{lm}f_{lm}$}, where $f_{lm}$ is a function of~$l,m$ that we will compute shortly. With this rescaling, Eq.~\eqref{eqn:cont_frac} becomes
\begin{equation}
\tilde{x}_{lm} = \frac{f_{lm}f_{l+1,m}B^{-}_{lm}/B^+_{lm}}{2\alpha_lf_{l+1,m}/B^+_{lm} - \tilde{x}_{l+1,m}}.
\end{equation}
We then choose $f_{lm}$ such that
\begin{equation}\label{eqn:defn_flm}
f_{lm}f_{l+1,m} = \frac{B^+_{lm}}{B^-_{lm}} 
\end{equation}
to move all of the $l$ dependence onto the $\alpha_l$. Then,
\begin{equation}
\tilde{x}_{lm} = \frac{1}{2\tilde{\alpha}_{lm}-\tilde{x}_{l+1,m}}, \ \tilde{\alpha}_{lm} = C_{lm}\left(\frac{\omega}{v_Fq}+i\frac{\gamma_l}{v_Fq}\right)\label{eqn:tild_x},
\end{equation}
where
\begin{equation}\label{eq:defClm}
    C_{lm} = f_{l+1,m}/B^+_{lm} = 1/(f_{lm}B^-_{lm}) .
\end{equation}
We determine the structure of $C_{lm}$ and its large-$l$ expansion in Appendix~\ref{app:calc_Clm}. Except for an overall prefactor $b_l$, the coefficients $C_{lm}$ are purely numerical coefficients that are independent of the Landau parameters as well as the relaxation rates.

\subsection{The conductivity in terms of $\tilde{x}_{lm}$}

Within linear response, the nonequilibrium distribution induced by the external electric field is proportional to the applied field, which allows us to determine the conductivity in terms of our solutions of Eq.~\eqref{eqn:Boltzmann_tower}.

The expression for the charge current within kinetic theory is~\cite{nozieres_theory_1999}
\begin{equation}
\bf{j} = - N_f eT\int \frac{\rm{d}^3\bf{p}}{(2\pi)^3}\bf{v}_{\bf{p}} \left(-\frac{\partial f_0}{\partial \varepsilon_{\bf{p}}}\right)\overline{\psi}_{\bf{p}},
\end{equation}
where we choose the convention \mbox{$e>0$} and include a factor of $N_f$ for the flavor number.
Inserting Eq.~\eqref{eqn:sph_decomp_Landau} into the expression for the current and expressing $\bf{v}_{\bf{p}}$ in spherical harmonics (cf.~Appendix~\ref{app:identities}) gives
\begin{align}
\bf{j} =& \  -\frac{eT\nu_0v_F}{\sqrt{24\pi}} \left(1+\frac{F_1}{3}\right)
\begin{pmatrix}
\psi_{1,-1} - \psi_{11} \\[0.5ex]
-i(\psi_{11} + \psi_{1,-1}) \\[0.5ex] \sqrt{2}\psi_{10}
\end{pmatrix} \label{eqn:current_psi},
\end{align}
where $v_F$ is the Fermi velocity and we define the total density of states
\begin{equation}
\nu_0 = \frac{N_f m^*}{2\pi^2\hbar^3}\sqrt{2m^*\varepsilon_F} = \frac{3}{2}\frac{n}{\varepsilon_F},
\end{equation}
where $n$ is the total density (including flavor degeneracy) in three dimensions. Note that the current only involves combinations of the \mbox{$l=1$} moments of the Fermi surface deformation, which describe a linear displacement of the Fermi surface. Finally, in our geometry \mbox{$j_x = \sigma_\perp E_x$}, \mbox{$j_y=\sigma_H E_y$}, and \mbox{$j_z = \sigma_{\parallel}E_z$} allowing us to easily determine the conductivity.

Our solution of the Boltzmann equation~\eqref{eqn:Boltzmann_tower} relied on transforming to the $x_{lm}$ variables of Eq.~\eqref{eqn:xlm}. We now want to express the current in terms of the electric field and these variables. We first note that since \mbox{$B^{\pm}_{l,-m} = B^{\pm}_{lm}$} then Eq.~\eqref{eqn:Boltzmann_tower} implies \mbox{$x_{l,-m} = x_{l,m}$} for \mbox{$l\geq 2$}. Then, in terms of \mbox{$x_{2,\pm 1}$}, Eq.~\eqref{eqn:Boltzmann_tower} for \mbox{$l=1, m=\pm 1$} is given by
\begin{equation}\label{eqn:psi11_determined}
\left(-\omega - i\gamma_i + B^+_{1,\pm 1}x_{2,\pm 1}\right) \psi_{1,\pm 1} = \pm i \sqrt{\frac{2\pi}{3}}e\beta E_0v_F\sin\theta_E.
\end{equation}
Using the fact that~\mbox{$x_{2,-1}=x_{21}$} and Eq.~\eqref{eqn:current_psi}, we obtain
\begin{align}
\sigma_\perp(\omega,q) =& \  \frac{i}{3}\left(1+\frac{F_1}{3}\right)\frac{\nu_0 v_F^2 e^2}{\omega + i\gamma_i - v_Fq\tilde{x}_{21}/2C_{11}}, \label{eqn:stat_cond}\\[1ex]
\sigma_H(\omega,q) =& \  0.
\end{align}
The latter equation makes physical sense: Since there is no time reversal breaking in our setup, there should be no Hall response. Thus, since the electric field is in the $xz$ plane, there cannot be any current in the $y$ direction by symmetry.

Solving for $\sigma_{\parallel}$ requires solving Eq.~\eqref{eqn:Boltzmann_tower} in terms of $x_{20}$, which in turns requires solving a matrix equation in terms of both $\psi_{00}$ and $\psi_{10}$. We obtain
\begin{equation}
\psi_{10} = \frac{i\sqrt{4\pi}e\beta E_0v_F \omega\cos\theta_E/\sqrt{3}}{\omega(\omega + i\gamma_i - v_Fq\tilde{x}_{20}/2C_{10})-c_1^2q^2},
\end{equation}
where
\begin{equation}
\frac{c_1^2}{v_F^2} = \frac{1}{3}\left(1+\frac{F_1}{3}\right)\left(1+F_0\right)
\end{equation}
is the speed of first sound in a 3D Fermi liquid~\cite{nozieres_theory_1999}. From this, we find the longitudinal conductivity in terms of $\tilde{x}_{20}$,
\begin{align}
\sigma_{\parallel}(\omega, q) &= \  \frac{i\omega e^2\nu_0 v_F^2(1+F_1/3)/3}{\omega(\omega + i\gamma_i) -c_1^2q^2 - v_Fq\omega \tilde{x}_{20}/2C_{10}}.
\end{align}

\subsection{Static conductivity}

We first investigate the static conductivity in the absence of impurity scattering, \mbox{$\gamma_i = 0$}. Equation~\eqref{eqn:stat_cond} reveals that this is given by
\begin{equation}
\sigma_\perp(q) = -0.78\frac{i\nu_0 v_Fe^2}{q\tilde{x}_{21}(0,q)},
\end{equation}
where we inserted the numerical value for $(1+F_1/3)C_{11}$. Further, Eq.~\eqref{eqn:tild_x} reveals that $\tilde{x}_{21}(0,q)$ is found by numerically solving the equation
\begin{equation}
\tilde{x}_{lm}(0,q) = \frac{1}{2iC_{lm}\gamma_l/v_Fq - \tilde{x}_{l+1,m}(0,q)}.
\end{equation}
We solve this using Lentz's algorithm, which gives the results shown in Fig.~\ref{fig:static_cond}. The physical interpretation of our results and their experimental consequences are discussed in Sec.~\ref{sec:methods_results}.

In Sec.~\ref{sec:methods_results}, we also plot the deformations of the Fermi surface in various limits. To do this, we note that from the definition of $x_{lm}$ in Eq.~\eqref{eqn:defn_xlm}
\begin{equation} \label{eqn:defn_psi_x}
\psi_{l,\pm 1} = x_{l,\pm 1}x_{l-1,\pm 1}\cdots x_{2,\pm 1}\psi_{1,\pm 1}.
\end{equation}
Thus, by solving for $\tilde{x}_{lm} = x_{lm}f_{lm}$ for all \mbox{$l\geq 2$} and using Eq.~\eqref{eqn:psi11_determined} to determine $\psi_{1,\pm 1}$, the deformation of all angular modes can be determined.

\subsection{Collective modes}

In this section, we would like to see how much about the collective modes of the transverse conductivity can be understood analytically. In order to do this, we take the scattering ansatz of Sec.~\ref{sec:methods_results}. Recall that with this ansatz we assumed \mbox{$\gamma_l = \gamma$} for \mbox{$l\geq 2$} and even, \mbox{$\gamma_l = \gamma'$} for \mbox{$l\geq 3$} and odd, and \mbox{$\gamma_1=\gamma_i$}, \mbox{$\gamma_0=0$}.
This allows us to extract the essential physics of the even-odd effect in 3D without committing to a particular form of the interelectron interaction $V$.

Before investigating the conductivity, we first return to Eq.~\eqref{eqn:tild_x} to analytically obtain a power series for $\tilde{x}_{21}(\omega,q)$ and $\tilde{x}_{20}(\omega,q)$ with this scattering ansatz. We have that 
\begin{equation}
\tilde{\alpha}_{lm} = C_{lm}\begin{cases} (\omega + i\gamma)/v_Fq &\mbox{if } l \text{ even, } l\geq 2 \\[1ex]
(\omega + i\gamma')/v_Fq &\mbox{if } l \text{ odd, } l\geq 3\end{cases}.
\end{equation}
We will further assume that \mbox{$F_l =0 $} for \mbox{$l\geq 2$}, as done in Ref.~\cite{hofmann_collective_2022}. Since we are only considering \mbox{$C_{lm}$ with $l\geq 2$}, this implies that we can set all of the $b_l$ [cf.~Eq.~\eqref{eqn:defn_Blm}] factors equal to $1$. Then, \mbox{$C_{lm}=1+{\it O}(l^{-2})$}, which allows us to see that $\tilde{\alpha}_{lm}-\tilde{\alpha}_{l+2,m} = {\it O}(l^{-3})$ in the limit of large $l$. 

We assume \mbox{$\tilde{x}_{lm} - \tilde{x}_{l+2,m} = {\it O}(l^{-3})$} as well and proceed to verify that this is a self-consistent assumption. This assumption allows us to perform a power series expansion in $1/l$ to solve the continued fraction. We write
\begin{equation}
\tilde{x}_{lm}(\omega,q) = y^{(0)}_m(\omega,q) + \frac{y^{(2)}_m(\omega,q)}{l^2} + \cdots.
\end{equation}
Then clearly
\begin{equation}
\tilde{x}_{lm} - \tilde{x}_{l+2,m} = \frac{4y^{(2)}_m(\omega,q)}{l^3}+\cdots,
\end{equation}
so this is consistent with our assumption that $\tilde{x}_{lm} - \tilde{x}_{l+2,m} = {\it O}(l^{-3})$. We can then solve the expression
\begin{equation}
\tilde{x}_{lm}(\omega,q) = \frac{1}{2C_{lm}\alpha - \frac{1}{2C_{l+1,m}\alpha' - \tilde{x}_{l+2,m}(\omega,q)}}
\end{equation}
order by order in $1/l$, where $\alpha = (\omega/v_Fq + i \gamma/v_Fq)$ and $\alpha' = (\omega/v_Fq + i \gamma'/v_Fq)$. Note that here we have taken $l$ to be even. When the power series expansion is solved, we find that
\begin{align}\label{eqn:x_tild_solution}
&\tilde{x}_{lm}(\omega,q) = \alpha'\biggl(1\pm \sqrt{1 - \frac{1}{\alpha\alpha'}}\biggr)\\
&\times \biggl(1 \pm \frac{4m^2-1}{8l^2} \frac{1}{\sqrt{1-\frac{1}{\alpha\alpha'}}}\biggl[1 \pm  \frac{1}{l}\frac{1}{\sqrt{1-\frac{1}{\alpha\alpha'}}}\biggr] \biggr) \nonumber,
\end{align}
to the third order in $1/l$. We note that this result is again consistent with \mbox{$\tilde{x}_{lm} - \tilde{x}_{l+2,m} = {\it O}(l^{-3})$}. So, our assumption was self-consistent.

We have written the power series as a nested series of parentheses to make it easy to compare the ratios of each successive term to one another. In order for this to be convergent, we need the coefficients of each successive term to remain finite. Inspecting these coefficients, we can see that the power series will converge as long as we do not have \mbox{$\alpha\alpha'\approx 1$}. Provided this requirement is satisfied, all of the additional numerical factors that emerged in 3D will not substantially alter the physics found in 2D. This is because the ${\it O}(l^0)$ term for $\tilde{x}_{lm}$ is precisely what was found in 2D~\cite{hofmann_collective_2022}.

Having found an analytical expression for $\tilde{x}_{lm}$ in terms of a power series in $1/l$, we now return to determine the collective modes of the transverse conductivity. The additional numerical factors in 3D make it very challenging to solve these equations analytically, so we will instead use what is known about the collective mode structure in 2D to focus on two specific limits. We first focus on the limit \mbox{$\gamma_i \ll \gamma' \ll v_Fq\ll \gamma$} and \mbox{$(q\xi)^2 \gg 1$}, where \mbox{$\xi = v_F/\sqrt{\gamma\gamma'}$} is a length scale. In this limit in 2D, it was found that the transverse conductivity exhibited a pole at \mbox{$\omega = -i\gamma' + {\it O}(\gamma\gamma'/(v_Fq)^2)$}~\cite{hofmann_collective_2022}. Thus, the collective mode of the transverse conductivity displayed signatures of the odd-parity scattering rate in this limit. We verify that generically, i.e., as long as \mbox{$F_1 > -3$}, the 3D Fermi liquid will also exhibit a collective mode with \mbox{$\omega = -i\gamma'$}.

To see this, let
\begin{equation}
\omega(q) = -i\gamma'-iA\frac{(\gamma' - \gamma_i)^2\gamma}{(v_Fq)^2} + {\it O}((q\xi)^{-3}),
\end{equation}
where $A$ is a dimensionless constant that will be chosen to give rise to a pole in transverse conductivity. We can evaluate \mbox{$\tilde{x}_{lm}(\omega(q),q)$} for this solution. Equation~\eqref{eqn:xlm} reveals that the factor \mbox{$(1-1/\alpha\alpha')^{-1/2}$} will scale like $(q\xi)^{-2}$. Thus, we see from Eq.~\eqref{eqn:xlm} that the $1/l$ corrections will be further suppressed by this factor. This allows us to see that for all even values of $l$ and all values of $m$ in this limit
\begin{equation}\label{eqn:tild_xlm_evenl}
\tilde{x}_{lm}(\omega(q), q) = -i\sqrt{A} \frac{\gamma'-\gamma_i}{v_Fq} + {\it O}((q\xi)^{-2}).
\end{equation}

A pole of the transverse conductivity satisfies the equation
\begin{equation}\label{eqn:pole_transverse}
0 = \omega(q) + i\gamma_i - \frac{v_Fq}{2C_{11}}\tilde{x}_{21}(\omega(q),q).
\end{equation}
Plugging in the above, we see that indeed $\omega(q)$ will be a solution to this equation as long as \mbox{$A = 4C_{11}^2$}. Using the solution of $C_{lm}$ discussed in Appendix~\ref{app:calc_Clm}, we see that this implies \mbox{$A\sim 5.5/(1+F_1/3)^2$}. Thus, as long as $F_1$ is not too close to $-3$, then $A$ will be on the order of $1$ or smaller, and this solution will be valid. Since \mbox{$F_1=-3$} indicates an instability \cite{nozieres_theory_1999} this condition will be satisfied in a generic 3D Fermi liquid. We conclude that just as in 2D, in the limit \mbox{$\gamma_i \ll \gamma' \ll v_Fq\ll \gamma$} and \mbox{$(q\xi)^2 \gg 1$}, we will have \mbox{$\omega=-i\gamma'$} as a pole of the transverse conductivity, serving as a way to probe the odd-parity mode scattering rates.

We can now produce the Fermi surface deformation corresponding to this collective mode.
If we choose a transverse electric field, then $\psi_{1,\pm 1}$ is proportional to the transverse conductivity, which diverges at the collective mode frequency. However, we can rescale all $\psi_{l,\pm 1}$ by this overall factor to accurately represent the shape of the collective mode. Then, working in the limit \mbox{$\gamma_i \ll \gamma' \ll v_Fq\ll \gamma$} and \mbox{$(q\xi)^2 \gg 1$} and using Eq.~\eqref{eqn:defn_xlm} reveals that in this limit the magnitude of $\psi_{l,\pm 1}$ for all odd $l$ will be equal, while it will be zero for all even $l$. We see that all odd-parity modes, which scatter with the frequency of the collective excitation, are excited in this limit.

As a contrast, we can also probe the limit $\gamma_i \ll v_Fq\ll \gamma' \ll \gamma$. Motivated by the collective mode of this limit in 2D \cite{hofmann_collective_2022}, we anticipate a collective mode with frequency
\begin{equation}
\omega(q) = -i\gamma_i - i A\frac{(v_Fq)^2}{\gamma},
\end{equation}
and will choose $A$ self-consistently in order to ensure this is a collective mode. Inserting this into Eq.~\eqref{eqn:xlm} reveals that each $1/l$ term is accompanied by a factor of $(1-1/A)^{-1/2}$. We will assume for now that $A\gg 1$, i.e., this term is of order $1$, and later confirm self-consistency. Then, 
\begin{equation}
\tilde{x}_{21}(\omega(q),q) \sim -iA\frac{v_Fq}{\gamma}\left(1\pm \sqrt{1-1/A}\right),
\end{equation}
where the coefficient of proportionality comes from completing the sum over $l$ in Eq.~\eqref{eqn:xlm}. We will not determine this since it will not affect the physics of the answer. Inserting this into Eq.~\eqref{eqn:pole_transverse} and solving in the limit of $A\gg 1$ will indeed give a solution, where \mbox{$A\sim 1+F_1/3$}. Thus, if we assume \mbox{$F_1\gg 1$}, this will be a self-consistent solution with
\begin{equation}
\omega(q) = -i\gamma_i - i B\left(1+\frac{F_1}{3}\right)\frac{(v_Fq)^2}{\gamma},
\end{equation}
where $B$ is a numerical coefficient that does not depend on the details of the interaction. Thus, we find a diffusive hydrodynamic mode, which does not depend on the odd-parity scattering rates.

Indeed, we can use the same logic as above to work out the Fermi surface deformation of the above collective mode. When this is done, we find that only the current zero mode, which does not decay, will contribute to the deformation. Thus, the Fermi surface deformation of this collective mode will not involve any contribution from modes that decay with the odd-parity scattering rate.

We conclude that the limit $\gamma_i \ll \gamma' \ll v_Fq\ll \gamma$ and $(q\xi)^2 \gg 1$ is an ideal limit for obtaining signatures of odd-parity scattering in the collective modes of the transverse conductivity.

\section{Conclusion}
\label{sec:conc_outlook} 

In this paper, we have investigated the decay rate of quasiparticle perturbations in a three-dimensional isotropic Fermi liquid. Contrary to the conventional wisdom that the tomographic effect relies on two-dimensional scattering kinematics, we reveal a tomographic effect in this three-dimensional setting, which is tunable by interactions. These results highlight that the separation between even- and odd-parity scattering rates is much more prevalent than assumed. Additionally, as highlighted in the introduction, the large Fermi temperatures in three dimensions may make the 3D version of the tomographic effect more straightforward to observe.

There are a number of promising directions for future research: First, the full collision integral could be evaluated, rather than the lowest order correction, as done in 2D in Refs.~\cite{hofmann_anomalously_2023, nilsson_nonequilibrium_2024}. This would allow for unambiguous identification of the scaling of decay rates with respect to a variety of experimentally relevant parameters, e.g., $r_s$, the ratio of Coulomb to kinetic energy. Another interesting avenue for exploration is whether including phonon scattering in the effective electron-electron interaction might give a possible route to realizing tunable long-range scattering in a realistic material. As discussed earlier, this would be expected to enhance the even-odd effect in 3D in a tunable way. Further experimental signatures of the even-odd effect in three dimensions should also be investigated. One possible signature could be obtained from further analysis of collective modes with realistic scattering rates (including those with long-range scattering). Additionally, the skin effect could be studied as a way to probe the transverse conductivity, as suggested in 2D by Refs.~\cite{baker_non-local_2023, valentinis_kinetic_2023, baker_nonlocal_2024}. Finally, the behavior of both the optical and static transverse conductivity in the presence of a magnetic field~\cite{rostami25,benshachar25a,benshachar25b} could be studied. As in 2D, the magnetic field would likely suppress the tomographic effect, allowing for a tunable comparison of even-odd scattering rates in a single material.

Finally, it would be interesting to study what effect dropping the assumption of isotropy has. We speculate that the even-odd effect may be further enhanced in an anisotropic Fermi liquid. For example, a quasi-two-dimensional Fermi surface, e.g., as found in some cuprates~\cite{fang_fermi_2022, musser_interpreting_2022}, is likely to have enhanced even-odd separation due to being closer to the two-dimensional limit. A systematic study of this limit might involve invoking the techniques of Refs.~\cite{hofmann_anomalously_2023, nilsson_nonequilibrium_2024} to exactly diagonalize the collision integral, as analytically evaluating the collision integral in the anisotropic limit is likely to prove challenging.

As all of these future directions highlight, the extension of tomographic effects to three dimensions provides an exciting and unexpected frontier for future research in the experimentally relevant setting of three-dimensional Fermi liquids.

\acknowledgments{
S.M.~and S.D.S.~were supported by the Laboratory for Physical Sciences. J.H.~was supported by Vetenskapsr\aa det (Grants No.~2020-04239 and No.~2024-04485), the Olle Engkvist Foundation  (Grant No.~233-0339), the Knut and Alice Wallenberg Foundation (Grant No.~KAW 2024.0129), and Nordita.
}

\appendix

\section{Useful definitions and identities}
\label{app:identities}

In this appendix, we collect identities for the spherical harmonics that are used in various places in the text, where we follow standard conventions in our definitions~\cite{nozieres_theory_1999,gottfriedyan03,arfken_mathematical_2013}. The spherical harmonics are defined as
\begin{equation} \label{eqn:defn_Ylm}
    Y_{lm}(\theta,\varphi) = (-1)^m \sqrt{\frac{2l+1}{4\pi} \frac{(l-m)!}{(l+m)!}}P^m_l(\cos\theta)e^{im\varphi},
\end{equation}
where $P^m_l$ is an associated Legendre polynomial. Their normalization is such that
\begin{equation}
\int \mathrm{d}\Omega \ Y_{lm}(\theta,\varphi)Y^*_{l'm'}(\theta,\varphi) = \delta_{ll'}\delta_{mm'},
\end{equation}
where $\mathrm{d}\Omega = \sin\theta \mathrm{d}\theta\mathrm{d}\varphi$ is the solid angle.

The first important relation is the addition theorem for spherical harmonics \cite{arfken_mathematical_2013}, which is used to express $\overline{\psi}_{\bf{p}}$ discussed in Sec.~\ref{sec:exp_sign} in the spherical harmonic basis as written in Eq.~\eqref{eqn:sph_decomp_Landau}. If $\xi$ is the angle between two solid angles $(\theta,\varphi)$ and $(\theta',\varphi')$ and $P_l$ is the $l$th Legendre polynomial, then 
\begin{equation}
P_l(\cos\xi) = \frac{4\pi}{2l+1}\sum_{m=-l}^{l}Y_{lm}(\theta,\varphi)Y^*_{lm}(\theta',\varphi').
\end{equation}
This is relevant to express the Fermi liquid correction to the quasiparticle energy: Since the Landau function $F$ is a function of $\cos\xi$, it can be decomposed into a sum of Legendre polynomials as 
\begin{equation}
F(\xi) = \sum_{l=0}^{\infty}F_lP_l(\cos \xi),
\end{equation}
while $\psi(\theta,\varphi)$ admits a decomposition into spherical harmonics as
\begin{equation}
\psi(\theta,\varphi) = \sum_{l=0}^{\infty}\sum_{m=-l}^{l}\delta \psi_{lm} Y_{lm}(\theta,\varphi) .
\end{equation}
This implies the following simplified form for the mean-field contribution to the quasiparticle energy:
\begin{equation}
\int \frac{\mathrm{d}\Omega'}{4\pi} \ F(\xi) \psi(\theta',\varphi') = \sum_{l=0}^{\infty}\frac{F_l}{2l+1}\sum_{m=-l}^{l}\psi_{lm}Y_{lm}(\theta,\varphi) , \label{eqn:int_decomp}
\end{equation}
which follows from the orthogonality properties of the spherical harmonics. Since $\overline{\psi}_{\bf{p}}$ is defined as \cite{nozieres_theory_1999}
\begin{equation}
 \overline{\psi}_{\bf{p}} = \psi_{\bf{p}} + \int \frac{\mathrm{d}\Omega'}{4\pi} \ F(\xi) \psi_{\bf{p}'},
\end{equation}
using~\eqref{eqn:int_decomp} in the above  gives  Eq.~\eqref{eqn:sph_decomp_Landau}.

Another relation we will need involves decomposing $Y_{lm}(\theta,\varphi)\cos\theta$ as a sum of two other spherical harmonics. One can use Eq.~\eqref{eqn:defn_Ylm} and the relation \cite{arfken_mathematical_2013}
\begin{equation}
xP^m_l(x) = \frac{l-m+1}{2l+1}P^m_{l+1}(x) + \frac{l+m}{2l+1}P^m_{l-1}(x),
\end{equation}
to see that
\begin{align}
Y_{lm}(\theta,\varphi)\cos\theta =& \ \sqrt{\frac{(l+1)^2-m^2}{4(l+1)^2-1}}Y_{l+1,m}(\theta,\varphi)\nonumber \\
&+ \sqrt{\frac{l^2-m^2}{4l^2-1}}Y_{l-1,m}(\theta,\varphi) \label{eqn:cos_decomp}.
\end{align}
Note that different values of $m$ remain decoupled. This is because the $\cos\theta$ term is symmetric about rotations around the $z$ axis and will thus not couple modes with different values of $m$.

Finally, in order to obtain Eq.~\eqref{eqn:Boltzmann_tower}, it is necessary to express $\cos\theta,\sin\theta\cos\varphi,$ and $\sin\theta\sin\varphi$ in terms of spherical harmonics. From Ref.~\cite{arfken_mathematical_2013}, we have
\begin{align}
\cos\theta &= \sqrt{\frac{4\pi}{3}}Y_{10}(\theta,\varphi), \\
\sin\theta\cos\varphi &= \sqrt{\frac{2\pi}{3}}\left(Y_{1,-1}(\theta,\varphi)-Y_{1,1}(\theta,\varphi)\right), \label{eqn:source_terms}
\end{align}
and
\begin{equation}
\sin\theta\sin\varphi = \sqrt{\frac{2\pi}{3}}i\left(Y_{1,-1}(\theta,\varphi)+Y_{1,1}(\theta,\varphi)\right) .
\end{equation}

\section{Orthonormal basis decomposition}\label{app:orth_basis_decomp}

In this appendix, we determine the  orthonormal basis for radial deformations of the quasiparticle deformations using the inner product defined in Eq.~\eqref{eqn:defn_inner_prod}. While we make use of the lowest order basis function in the main text, any future numerical solution of the collision integral, as done in 2D in Refs.~\cite{hofmann_anomalously_2023, nilsson_nonequilibrium_2024}, will require the higher-order basis functions that we establish here.

As discussed in the main text, the symmetry of the linearized collision integral under a global rotation implies that deformations are parametrized by a separation of variables in terms of spherical harmonics and a radial deformation. We choose the normalization of Eq.~\eqref{eqn:defn_Ylm} and 
evaluate~\mbox{$\langle u_{n'}Y_{l',m'}|u_nY_{lm}\rangle$} by first integrating out the angular dependence. We then make a change of variables to~\mbox{$w = \beta(\varepsilon_{\bf{p}}-\mu)$}, which gives
\begin{align}
&\langle u_{n'}Y_{l',m'}|u_nY_{lm}\rangle \nonumber \\
&\quad =\frac{\delta_{l'l}\delta_{m'm}}{\sqrt{\mu}}\int_{-\beta \mu}^{\infty} \frac{u^{*\prime}_{n'}(w)u_n(w)\sqrt{Tw+\mu}\ {\rm d}w}{4\cosh^2(w/2)} .
\end{align}
To ensure orthonormality, we need the integrand to evaluate to~$\delta_{n'n}$. 

We note that the~$\cosh^2(w/2)$ factor in the denominator ensures that~$w$ will be of order $1$ in the integrand. Thus, at low temperatures, we expand~\mbox{$\sqrt{Tw+\mu}$}
as~\mbox{$\sqrt{\mu} + {\it O}(T/T_F)$}. Hence, we choose~$u_n$ to satisfy
\begin{equation}
\delta_{n'n} = \int_{-\infty}^{\infty}\frac{u^{*\prime}_{n'}(w)u_n(w)\ {\rm d}w}{4\cosh^2(w/2)}.
\end{equation}
This is precisely the same condition as in two dimensions, where the lowest order basis functions were found to be \cite{hofmann_collective_2022,nilsson_nonequilibrium_2024}
\begin{align}
u_0(w) =& \  1,\\
u_1(w) =& \  \frac{\sqrt{3}}{\pi} w, \label{eq:linearbasis}
\end{align}
and so on. We will only make use of the leading order expression $u_0(w)=1$.

\section{Integrating out a global rotation}\label{app:int_out_rot}

In this appendix, we explicitly verify that the linearized collision integral~\eqref{eqn:lin_coll_mat} is diagonal in the angular basis~$l,m$. Of course, as discussed in Sec.~\ref{sec:coll_int_three}, this follows on general grounds from the Wigner-Eckart theorem~\cite{arfken_mathematical_2013}. Here, we verify this fact by integrating out a global rotation in the collision integral. The resulting explicit expression proves useful when calculating the scattering rates.

We begin by rewriting the matrix element of the linearized collision integral
\begin{align}
&{\cal L}_{n'l'm',nlm} = \  \langle u_{n'}Y_{l',m'}|{\cal L}|u_nY_{lm}\rangle \nonumber \\
&= \  \frac{(\pi \lambda_T)^2}{p_F} \iiiint \frac{d(\tilde{{\bf p}}_1,\tilde{{\bf p}}_2,\tilde{{\bf p}}_1',\tilde{{\bf p}}_2')}{(2\pi)^{12}}
W(\tilde{\bf{p}}_1', \tilde{\bf{p}}_2'|\tilde{\bf{p}}_1,\tilde{\bf{p}}_2)\nonumber\\
&\times F_{\tilde{1}\tilde{2}\tilde{1}'\tilde{2}'} {\cal S}_{n'l'm'}^*(\tilde{\bf{p}}_1, \tilde{\bf{p}}_2, \tilde{\bf{p}}_1', \tilde{\bf{p}}_2') {\cal S}_{nlm}(\tilde{\bf{p}}_1, \tilde{\bf{p}}_2, \tilde{\bf{p}}_1', \tilde{\bf{p}}_2') \label{eqn:coll_mat_elem} ,
\end{align}
where $u_n$ are radial basis functions defined in Appendix~\ref{app:orth_basis_decomp}, $Y_{lm}$ are spherical harmonics, $F_{\tilde{1}\tilde{2}\tilde{1}'\tilde{2}'}$ is defined in Eq.~\eqref{eqn:fermi_func_prod}, and we abbreviate
\begin{align}
{\cal S}_{nlm}(\bf{p}_1,\bf{p}_2,\bf{p}_1',\bf{p}_2') =& \ u_{n}Y_{lm}(\bf{p}_1) + u_{n}Y_{lm}(\bf{p}_2)\nonumber \\
&- u_{n}Y_{lm}(\bf{p}_1') -  u_{n}Y_{lm}(\bf{p}_2') ,
\end{align}
where we have used our freedom to relabel  dummy variables with tildes.

To evaluate this expression, we work in a coordinate system $xyz$ that is invariant under rotations~\footnote{This is the convention when working with Euler angles~\cite{arfken_mathematical_2013}.}, and label the operator implementing an active rotation by~$\varphi$ about the \mbox{$n=x,y,z$}-axis (in the mathematically positive sense) $R_n(\varphi)$. We define
\begin{align}
\tilde{\bf{p}}_1 =& \  R_z(\varphi_1)R_y(\theta_1)R_z(\varphi_2) [p_1\hat{\bf{z}}] \nonumber\\
\tilde{\bf{p}}_2 =& \  R_z(\varphi_1)R_y(\theta_1)R_z(\varphi_2)[\bf{p}_2^{xz}] \nonumber \\
\tilde{\bf{p}}_1' =& \  R_z(\varphi_1)R_y(\theta_1)R_z(\varphi_2)\bf{p}_1'\nonumber \\
\tilde{\bf{p}}_2' =& \  R_z(\varphi_1)R_y(\theta_1)R_z(\varphi_2)\bf{p}_2' ,
\end{align}
where $\bf{p}_2^{xz} = p_2\sin\theta_2 \hat{\bf{x}} + p_2\cos\theta_2\hat{\bf{z}}$ lies in the $xz$ plane, $\theta_1,\varphi_1$ are the angular coordinates of $\bf{p}_1$, and $\varphi_2$ is the azimuthal angle of $\bf{p}_2$. It is straightforward to check that the Jacobian of this transformation is equal to unity. This gives
\begin{align}
&{\cal L}_{n'l'm',nlm} = \ \frac{(\pi \lambda_T)^2}{p_F}\iiiint \frac{d({{\bf p}}_1,{{\bf p}}_2,{{\bf p}}_1',{{\bf p}}_2')}{(2\pi)^{12}} \nonumber \\
&\quad \times W(\bf{p}_1',\bf{p}_2'|p_1\hat{\bf{z}}, \bf{p}_2^{xz})F_{121'2'}\nonumber \\[1ex]
&\quad \times {\cal S}_{n'l'm'}^*(\tilde{\bf{p}}_1, \tilde{\bf{p}}_2, \tilde{\bf{p}}_1', \tilde{\bf{p}}_2') {\cal S}_{nlm}(\tilde{\bf{p}}_1, \tilde{\bf{p}}_2, \tilde{\bf{p}}_1', \tilde{\bf{p}}_2'), \nonumber 
\end{align}
where we have used rotation invariance to factor out the rotations from the $W$ and $F$ pieces of the integrand. 

Next, we would like to factor out the rotation from the ${\cal S}$ term. By the properties of the spherical harmonics under rotation~\cite{arfken_mathematical_2013}, we have for all $\bf{p}$
\begin{align}
&Y_{lm}\left[R_z(\varphi_1)R_y(\theta_1)R_z(\varphi_2)\bf{p}\right] \nonumber \\
&= \sum_{m'=-l}^{l}D^{l*}_{mm'}(\varphi_1, \theta_1, \varphi_2) Y_{l,m'}(\bf{p}) ,
\end{align}
where $D^l_{mm'}(\varphi_1, \theta_1, \varphi_2)$ is the Wigner D-matrix for the rotation $R_z(\varphi_1)R_y(\theta_1)R_z(\varphi_2)$.
We see that
\begin{align}
&{\cal S}_{n'l'm'}^*(\tilde{\bf{p}}_i) {\cal S}_{nlm}(\tilde{\bf{p}}_i) \nonumber \\
&= \  \sum_{m'''=-l'}^{l'}\sum_{m''=-l}^{l} D^{l'}_{m'm'''}(\varphi_1,\theta_1,\varphi_2) D^{l*}_{mm''}(\varphi_1,\theta_1,\varphi_2) \nonumber \\
&\quad \times {\cal S}_{n'l'm'''}^*(p_1\hat{\bf{z}}, \bf{p}_2^{xz}, \bf{p}_1', \bf{p}_2') {\cal S}_{nlm''}(p_1\hat{\bf{z}},  \bf{p}_2^{xz}, \bf{p}_1', \bf{p}_2') .
\end{align}
Having factored out the overall rotation, we note that $p_1\hat{\bf{z}},\bf{p}_2^{xz},\bf{p}_1',\bf{p}_2'$ all do not depend on $\varphi_1,\theta_1,\varphi_2$. Hence, we can freely integrate the D-matrices over these variables. This implies
\begin{widetext}
\begin{align}
{\cal L}_{n'l'm',nlm} &= \frac{(\pi \lambda_T)^2}{p_F}\int \frac{p_1^2p_2^2{\rm d}p_1 {\rm d}p_2 {\rm d}(\cos\theta_2) {\rm d} \bf{p}_1'{\rm d} \bf{p}_2'}{(2\pi)^{6}} W(\bf{p}_1',\bf{p}_2'|p_1\hat{\bf{z}}, \bf{p}_2^{xz})F_{121'2'}\nonumber \\
&\quad \times \sum_{m'''=-l'}^{l'}\sum_{m''=-l}^{l} \biggl[{\cal S}_{n'l'm'''}^*(p_1\hat{\bf{z}}, \bf{p}_2^{xz}, \bf{p}_1', \bf{p}_2') 
{\cal S}_{nlm''}(p_1\hat{\bf{z}},  \bf{p}_2^{xz}, \bf{p}_1', \bf{p}_2') \nonumber \\
&\quad\qquad\times \frac{1}{(2\pi)^6}\int_0^{2\pi}{\rm d}\varphi_1{\rm d}\varphi_2 \int_0^{\pi}{\rm d}(\cos\theta_1)
D^{l'}_{m'm'''}(\varphi_1,\theta_1,\varphi_2)D^{l*}_{mm''}(\varphi_1,\theta_1,\varphi_2)\biggr] .
\end{align}
We now use the orthogonality relations of the Wigner D-matrix~\cite{arfken_mathematical_2013}, which gives
\begin{align}
&{\cal L}_{n'l'm',nlm} = \  \delta_{l'l}\delta_{m'm}\frac{\lambda_T^2}{8\pi^2(2l+1)p_F} \nonumber \\
&\quad \times \sum_{m''=-l}^{l} \int \frac{p_1^2p_2^2{\rm d}p_1 {\rm d}p_2 {\rm d}(\cos\theta_2) {\rm d} \bf{p}_1'{\rm d} \bf{p}_2'}{(2\pi)^{6}} 
W(\bf{p}_1',\bf{p}_2'|p_1\hat{\bf{z}}, \bf{p}_2^{xz})F_{121'2'}
{\cal S}_{n'lm''}^*(p_1\hat{\bf{z}}, \bf{p}_2^{xz}, \bf{p}_1', \bf{p}_2') 
{\cal S}_{nlm''}(p_1\hat{\bf{z}},  \bf{p}_2^{xz}, \bf{p}_1', \bf{p}_2'). 
\end{align}
\end{widetext}
We have thus shown by an explicit calculation that the collision matrix is diagonal with respect to $l$ and $m$. In fact, the above result reveals that there is no dependence of the collision matrix element on~$m$ at all. As mentioned in the main text, this follows from an extension of the Wigner-Eckart theorem~\cite{arfken_mathematical_2013}.

\section{Calculating $({\cal L}_l)_{0,0}$}
\label{app:calc_gamml}

In this appendix, we calculate the leading order contribution to the scattering rates $\gamma_l$ as defined in Sec.~\ref{sec:coll_int_three}. In doing so, we use the results of Appendices~\ref{app:orth_basis_decomp} and~\ref{app:int_out_rot}. Note that while Sec.~\ref{sec:coll_int_three} summarizes some of the results of this appendix, this appendix is as self-contained as possible for ease of reading.

As a first step to evaluate the leading-order contribution to $\gamma_l$, we integrate over the delta function in Eq.~\eqref{eqn:Coulomb_scatt_mat} that enforces momentum conservation, which expresses both $\bf{p}_1'$ and $\bf{p}_2'$ in terms of the momentum transfer \mbox{$\bf{k} = \bf{p}_1-\bf{p}_1'=\bf{p}_2'-\bf{p}_2$}, displayed in Fig.~\ref{fig:2d_vs_3d}. When this is done we find that
\begin{align}
&({\cal L}_{l})_{0,0} = \ \frac{2\pi^3 \lambda_T^2}{\hbar p_F} \iiint \frac{d(\bf{p}_1,\bf{p}_1,\bf{k})}{(2\pi)^9} |V|^2F_{121'2'}\nonumber \\[1ex]
&\times |{\cal S}_{0lm',121'2'}|^2 \delta(\varepsilon_{{\bf p}_1} + \varepsilon_{{\bf p}_2} - \varepsilon_{{\bf p}_1-\bf{k}} - \varepsilon_{{\bf p}_2+\bf{k}})
\end{align}
with
\begin{align}
{\cal S}_{0lm,121'2'} &= Y_{lm}(\bf{p}_1) + Y_{lm}(\bf{p}_2) - Y_{lm}(\bf{p}_1') - Y_{lm}(\bf{p}_2') . 
\end{align}
As we will see, the $m'$ dependence will later vanish.

The next step is to transform to a convenient basis. To this end, we use the manipulations detailed in App.~\ref{app:int_out_rot} to rotate $\bf{k}$ to point along the $z$ axis and $\bf{p}_1$ to lie in the $xz$ plane. This integrates out all angular dependence of $\bf{k}$ and the~$\varphi$ dependence of $\bf{p}_1$. This gives
\begin{align}\label{eq:appDmatrixelement}
&({\cal L}_l)_{0,0} = \  \frac{\lambda_T^2}{4\pi \hbar p_F (2l+1)} \sum_{m}\int \frac{k^2p_1^2{\rm d}k{\rm d}p_1{\rm d}(\cos\theta_1) {\rm d}\bf{p}_2}{(2\pi)^3} \nonumber \\
&\times |V|^2 F_{121'2'}\delta(\varepsilon_{{\bf p}_1} + \varepsilon_{{\bf p}_2} - \varepsilon_{{\bf p}_1'} - \varepsilon_{{\bf p}_2'}) \nonumber 
\\
&\times \left|Y_{lm}(\theta_1,0) + Y_{lm}(\theta_2,\varphi_2) - Y_{lm}(\theta_1',0)-Y_{lm}(\theta_2',\varphi_2)\right|^2 .
\end{align}
Here, $\theta_1'$ is a function of $p_1,k$, and $\theta_1$ since \mbox{$\bf{p}_1'=\bf{p}_1-\bf{k}$}, and likewise for $\theta_2'$.

\section{Calculating $C_{lm}$}
\label{app:calc_Clm}

\begin{figure}[b!]
    \centering
    \includegraphics[width=0.95\columnwidth]{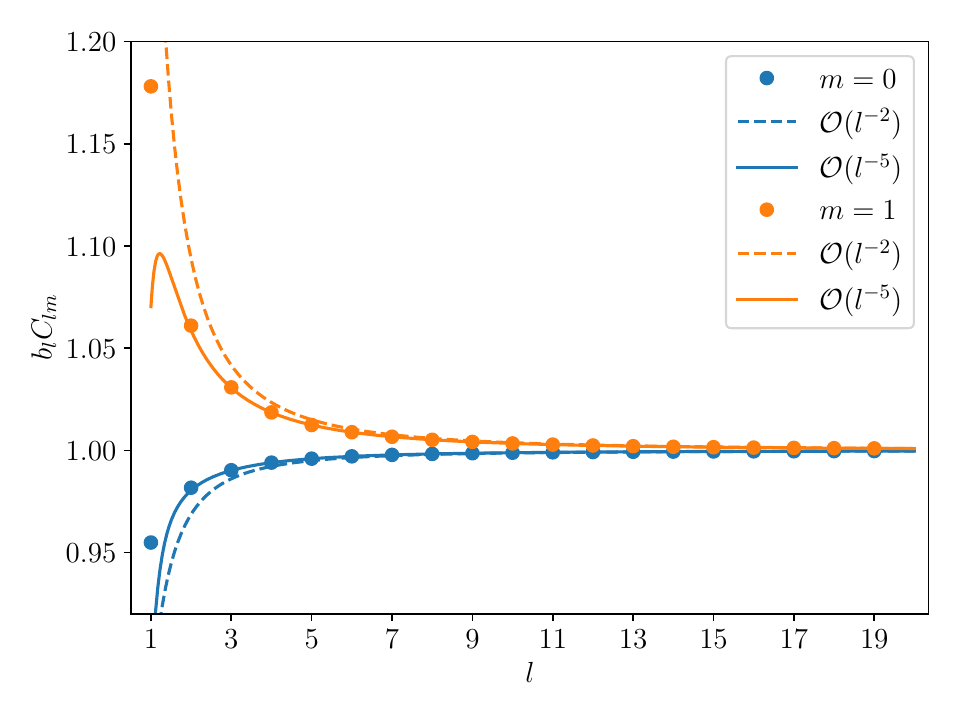}
    \caption{Scaling function~$b_lC_{lm}$ as a function of~$l$ for~$m=0,1$. The points are numerical values obtained using Eq.~\eqref{eqn:infinite_Blm} and the dotted (solid) lines are the series expansion up to $l^{-2}$ ($l^{-5}$) [Eq.~\eqref{eqn:series_Blm}]. We find close agreement for all~$l\geq 2$ plotted. The interpretation of this is discussed in the main text; we can see that in the static limit the $m=1$ scatterings are enhanced at small $l$, while they are suppressed at small $l$ when $m=0$. }
    \label{fig:scaling}
\end{figure}

In this appendix, we calculate the factors $f_{lm}$ defined in Eq.~\eqref{eqn:defn_flm} of Sec.~\ref{sec:exp_sign}, along with the coefficients~$C_{lm}$ defined in the same section. 

We first determine the value of~$f_{lm}$ using the boundary condition~\mbox{$\lim_{l\rightarrow \infty}f_{lm}=1$}. From Eq.~\eqref{eqn:defn_flm}, we note that
\begin{align}
f_{lm} =& \ \frac{B^{+}_{lm}}{B^-_{lm}}\frac{1}{f_{l+1,m}} = \cdots\\
=& \frac{\left(\prod_{k=0}^{\infty} B^+_{l+2k,m}\right)\left(\prod_{k=0}^{\infty} B^-_{l+1+2k,m}\right)}{\left(\prod_{k=0}^{\infty}B^-_{l+2k,m}\right)\left(\prod_{k=0}^{\infty} B^+_{l+1+2k,m}\right)} \nonumber .
\end{align}
Recalling the definition of~$B^{\pm}_{lm}$ in Eq.~\eqref{eqn:defn_Blm}, in the infinite product above, we can factor out all of the $b_l$ terms, and almost all of them will cancel out. The only remaining term is a factor of~\mbox{$b_l/b_{l-1}$}. The remaining infinite product is simplified by taking the logarithm to turn it into an infinite sum, which gives
\begin{align}
f_{lm} =& \  \frac{b_l}{B^-_{lm}}\exp\left[-\sum_{n=1}^{\infty}(-1)^n\log\left(\frac{(l+n)^2-m^2}{(l+n)^2-1/4}\right)\right] 
\end{align}
and, from Eq.~\eqref{eq:defClm},
\begin{align}
C_{lm} =& \  \frac{1}{b_l}\exp\left[\sum_{n=1}^{\infty}(-1)^n\log\left(\frac{(l+n)^2-m^2}{(l+n)^2-1/4}\right)\right] \label{eqn:infinite_Blm}.
\end{align}
Note that $f_{lm}$ is only defined for \mbox{$l>|m|$}, while $C_{lm}$ is defined for \mbox{$l\geq |m|$}. While this is not an issue as the spherical harmonic decomposition requires \mbox{$|m|\leq l$} in the first place, it implies that both~\mbox{$|f_{lm}-1|$} and~\mbox{$|C_{lm}-1|$} asymptote as~$l$ approaches~$m$.

These expressions converge quickly and are thus straightforward to calculate numerically. We show~$b_lC_{lm}$ in Fig.~\ref{fig:scaling}; this is a function of $l$ and $m$ only and does not depend on $\omega$ or any of the scattering rates. This means that the only physical parameter that~$C_{lm}$ depends on is~$b_l^{-1} = [1+F_l/(2l+1)]^{-1}$. 

We can gain additional analytical insight by rewriting the infinite sum in the limit~$l+n \gg 1,m$. There, we see that
\begin{align}
& \log(b_lC_{lm}) = \  \sum_{n=1}^{\infty}\sum_{k=1}^{\infty} \left(\frac{1-(2m)^{2k}}{4^kk}\right) \frac{(-1)^n}{(l+n)^{2k}} \nonumber \\
&= \sum_{k=1}^{\infty} \left(\frac{1-(2m)^{2k}}{16^kk}\right) \left[\zeta\left(2k,\frac{l+2}{2}\right) - \zeta\left(2k,\frac{l+1}{2}\right)\right] ,
\end{align}
where~$\zeta$ is the Hurwitz zeta function. The utility of doing this is that in the~\mbox{$l\gg 1$} limit, the~$k$th term of this new sum contributes only at ${\it O}(l^{-2k})$. This makes it easy to expand the sum in the large $l$ limit since only a finite number of terms are needed. When this is done, we obtain
\begin{align}
b_lC_{lm} =& \ 1 + \frac{4m^2-1}{8l^2}\left(1 - \frac{1}{l} + \frac{1+12m^2}{16l^2} + \frac{7-12m^2}{8l^3}\right)\nonumber \\
&+ {\it O}\left(\frac{1}{l^6}\right) . \label{eqn:series_Blm}
\end{align}
Figure~\ref{fig:scaling} shows this series up to the~$l^{-5}$ term compared to the exact numerical value~\eqref{eqn:infinite_Blm}, where we find excellent agreement for all values of~$l\geq 2$.

\bibliographystyle{apsrev4-1_custom}
\bibliography{integral.bib}

\end{document}